\documentclass[journal,draftcls,onecolumn,letterpaper,twoside,11pt]{IEEEtran}
\ifCLASSINFOpdf
\else
\fi
\usepackage{graphicx}
\usepackage{color}

\begin{document}
%
\title{Chaotic Method for Generating $q$-Gaussian Random Variables}
%
%
%

\author{Ken~Umeno,~Aki-Hiro~Sato
\thanks{Ken Umeno is with the Department of Applied
Mathematics and Physics, Graduate School of Informatics, Kyoto
University, Yoshida Honmachi, Sakyo-ku, Kyoto 606-8501 JAPAN
(e-mail:umeno.ken.8z@kyoto-u.ac.jp)} 
\thanks{Aki-Hiro Sato is with the Department of Applied
Mathematics and Physics, Graduate School of Informatics, Kyoto
University, Yoshida Honmachi, Sakyo-ku, Kyoto 606-8501 JAPAN
(e-mail:sato.akihiro.5m@kyoto-u.ac.jp)}
\thanks{Color versions of Figures 1--4 in this correspondence are available
online.}}

\maketitle

\begin{abstract}
This study proposes a pseudo random \textcolor{black}{number} generator
of $q$-Gaussian random variables for a range of $q$ values, $-\infty <
q < 3$, based on deterministic \textcolor{black}{chaotic} map dynamics. 
Our method consists of chaotic maps on the unit circle and
map dynamics based on \textcolor{black}{the piecewise linear map}. We perform the 
$q$-Gaussian random \textcolor{black}{number} generator for
several values of $q$ and conduct \textcolor{black}{both
Kolmogorov-Smirnov (KS) and Anderson-Darling (AD) tests. The
$q$-Gaussian samples generated by our proposed method pass the KS
test at more than 5\% significance level for values of $q$
ranging from -1.0 to 2.7, while they pass the AD test at more than 5\%
significance level for $q$ ranging from -1 to 2.4.}  
\end{abstract}


%
\IEEEpeerreviewmaketitle

\section{Introduction}
%
%
%
%
\IEEEPARstart{T}{he} $q$-Gaussian distribution\textcolor{black}{s} 
\textcolor{black}{have} been studied in a wide
variety of fields from natural sciences to social sciences. 
\textcolor{black}{They} have been applied in thermodynamics,
biology, economics, and quantum mechanics. The generating
mechanism is still an open question, but several mechanisms
that have been shown to produce $q$-Gaussian distributions
\textcolor{black}{are known, such as} multiplicative noise, weakly chaotic
dynamics, correlated anomalous diffusion, preferential growth of
networks, and asymptotically scale-invariant correlations~\cite{Tsallis}. 
\textcolor{black}{In the heavy-tail domain ($1 < q < 3$), the
  $q$-Gaussian distribution is equivalent to the Student's
  $t$-distribution.}

\textcolor{black}{In the context of finance, the $q$-Gaussian
  distribution ($1 < q < 3$) is referred to as a Student's $t$
  distribution~\cite{Student}. This is commonly used in finance and
  risk management, particularly to model conditional asset returns of
  which the tails are wider than those of normal distribution. The
  distribution is also known as Pearson Type-II (for compact support
  ($q<1$) and Type VII (infinite support ($q\geq 1$)~\cite{Pearson}.
  For example, Bollerslev used the Student's $t$ to model the
  distribution of the foreign exchange rate
  returns~\cite{Bollerslev}. Bening and Korolev provide an instance
  where the distribution is appropriate as a model, i.e. the case of
  random sample sizes~\cite{Bening}. Vignat and Plastino obtained
  similar results~\cite{Vignat}. Other work attempts to show the
  $q$-Gaussian distribution as an attractor in the context of
  dependent systems~\cite{Umarov}. Moreover, Umarov et
  al. consider a $q$-extension of $\alpha$-stable L\'evy
  distribution~\cite{Umarov:10}.} 

More recently, $q$-Gaussian distributions have been derived from the 
maximization of non-extensive entropy~\cite{Tsallis} and studied in the
context of the generalization of Gauss' Law of
Errors~\cite{Suyari}. $q$-Gaussian distributions can be derived from
an infinite normal mixture with an 
inverse gamma distribution. This concept is known as superstatistics in
non-equilibrium thermodynamics~\cite{Beck}. $q$-Gaussian distributions
also appear as unconditional distributions of multiplicative
stochastic differential equations~\cite{Sato}.

Recently, the generalized Box-Muller method \textcolor{black}{(GBMM)}
was proposed by Thisleton et al.~\cite{Thistleton:07}. Their method
uses transformation including the $q$-logarithmic, sine, and cosine
functions in terms of uniform random variables. Here, based on the
ergodic theory~\cite{Lasota} of dynamical 
systems, we propose a family of chaotic maps with an ergodic invariant
measure given by $q$-Gaussian density. Ulam and von Neumann considered
the logistic map $X_{n+1}=4X_{n}(1-X_{n})$ in the late 1940s, and found
its randomness~\cite{Ulam}. One of the authors
(K. Umeno) proposed chaotic mechanism to generate power-law random
variables\textcolor{black}{~\cite{Umeno:98}}. This method can generate
power-law random variables in the L\'evy stable regime from the
superposition of \textcolor{black}{the random variables}. One of the
authors (A.-H. Sato) also proposed multiplicative random processes to
generate power-law random variables~\cite{Takayasu}. Currently,
we can use the map dynamics to design random sequences
\textcolor{black}{with an explicit} ergodic \textcolor{black}{invariant}
measure more precisely~\cite{Umeno:97,Chen}.

In this article, we propose a method to generate $q$-Gaussian random 
variables based on deterministic map dynamics. Our method is based on
ergodic transformations on the unit circle and a map composed of the 
piecewise linear map with both the $q$-exponential and $q$-logarithmic
function. This method is a direct method different from
Ref.~\cite{Thistleton:07,Umeno:98} and can generate $q$-Gaussian
random variables for $-\infty < q < 3$ including
\textcolor{black}{infinite variance and infinite mean regimes.} We
generate $q$-Gaussian random variables for several cases of $q$, and
conduct statistical testing by means of analytical cumulative
distribution functions.

\section{Review of the generalized Box-Muller Method}
The zero-mean normal $q$-Gaussian distribution parameterized by $q$
is described as 
\begin{equation}
g(x;q) =
\left\{
\begin{array}{ll}
\frac{1}{B\Bigl(\frac{2-q}{1-q},\frac{1}{2}\Bigr)}\Bigl(\frac{1-q}{3-q}\Bigr)^{\frac{1}{2}}\Bigl(1\textcolor{black}{-\frac{1-q}{3-q}}x^2\Bigr)^{\frac{1}{1-q}}
 & (q < 1) \\
\frac{1}{\sqrt{2\pi}}\exp\Bigl(-\frac{x^2}{2}\Bigr) & (q=1) \\
 \frac{1}{B\Bigl(\frac{1}{1-{q}}-\frac{1}{2},\frac{1}{2}\Bigr)}\Bigl(\frac{{q}-1}{3-{q}}\Bigr)^{\frac{1}{2}}\Bigl(1\textcolor{black}{+\frac{q-1}{3-q}}x^2\Bigr)^{\frac{1}{1-q}}
  & (1 < q < 3)
\end{array}
\right.,
\label{eq:q-gauss-def}
\end{equation}
where $B(a,b)$ is the beta function, which is defined as
\begin{equation}
B(a,b) = \int_{0}^{1}t^{a-1}(1-t)^{b-1}\mbox{d}t.
\end{equation}
For $q < 1$ symmetric distributions with compact support ranging from
$-\sqrt{\frac{2}{1-q}}$ to $\sqrt{\frac{2}{1-q}}$ appear. Specifically,
the normalized Wigner distribution is obtained at $q = -1$. 
In the case of $1 < q < 3$, Equation \ref{eq:q-gauss-def} has
\textcolor{black}{heavy-tails and $g(x;q) \approx const. |x|^{\nu-1}$, where
$\nu=(3-q)/(q-1) > 0$ is related to the degree of freedom of the
Student's $t$-distribution. $\nu$ is coincident with the tail index
of the complementary cumulative distribution of $g(x;q)$.} This also
gives an existence condition in the heavy-tail regime of the
$q$-Gaussian distribution.

Firstly, let us start our discussion from the \textcolor{black}{GBMM} proposed by
Thistleton et al.~\cite{Thistleton:07}. \textcolor{black}{To introduce
their method to generate $q$-Gaussian random variable, we define a
$q$-analog of both exponential and logarithmic function.}

\textcolor{black}{
{\bf Definition 1.} 
Suppose the one-dimensional ordinary differential equation
\begin{equation}
\frac{\mbox{d}h}{\mbox{d}w} = h^{q}, \quad h(0) = 1.
\end{equation}
The solution is given as
\begin{equation}
h(w)=
\left\{
\begin{array}{ll}
\Bigl(1+(1-q)w\Bigr)^{\frac{1}{1-q}} & 1+(1-q)w > 0 \\
0 & \mbox{elsewhere} \\
\end{array}
\right..
\label{eq:q-exp}
\end{equation}
We call the solution $h(w)$ $q$-exponential function. Obviously, one has
\begin{equation}
\lim_{q \rightarrow 1} \Bigl(1+(1-q)w\Bigr)^{\frac{1}{1-q}} = e^{w}.
\end{equation}
}

\textcolor{black}{
{\bf Definition 2.}
We define the inverse function of Equation \ref{eq:q-exp}
\begin{equation}
\ln_q(w) = \frac{w^{1-q}-1}{1-q} \quad (w > 0),
\label{eq:q-log}
\end{equation}
which we call the $q$-logarithmic function. Clearly, we get
\begin{equation}
\lim_{q \rightarrow 1} \frac{w^{1-q}-1}{1-q} = \ln w.
\end{equation}
} 

\textcolor{black}{
{\bf Definition 3.}
The GBMM~\cite{Thistleton:07} is given by transformations from
i.i.d. uniform random variables $u_1$ and $u_2$ ranging from 0 to 1.
\begin{equation}
\left\{
\begin{array}{lll}
x &=& \sqrt{-2\ln_q(u_1)}\cos(2\pi u_2) \\
y &=& \sqrt{-2\ln_q(u_1)}\sin(2\pi u_2)
\end{array}
\right..
\label{eq:GBMM}
\end{equation}
}

\textcolor{black}{
{\bf Proposition 1.}
The joint probability density of $x$ and $y$ in Equation \ref{eq:GBMM}, 
is given by
\begin{equation}
p_{X,Y}(x,y) =
\frac{1}{2\pi} \exp_{2-1/q}\Bigl(-\frac{q}{2}(x^2+y^2)\Bigr).
\label{eq:2-q-gauss}
\end{equation}
}

\textcolor{black}{
{\bf Proof of Proposition 1.}
From Equation \ref{eq:GBMM}, we obtain
\begin{eqnarray}
\nonumber
p_{U_1}(u_1)p_{U_2}(u_2) &=& p_{X,Y}(x,y) \Bigl|\frac{\partial(x,y)}{\partial(u_1,u_2)} \Bigr|\\
\nonumber
1 &=& 2\pi p_{X,Y}(x,y) u_1^{-q} \\
\nonumber
p_{X,Y}(x,y) &=& \frac{1}{2\pi}\Bigl[\exp_q\bigl(-\frac{1}{2}(x^2+y^2)\bigr)\Bigr]^q \\
&=& \frac{1}{2\pi}\exp_{2-1/q}\Bigl(-\frac{q}{2}(x^2+y^2)\Bigr), 
\label{eq:2d-qgauss}
\end{eqnarray}
where we used the equality 
\begin{equation}
\bigl(\exp_q(x)\bigr)^q = \bigl(1+(1-q)x\bigr)^{\frac{q}{1-q}} =
\exp_{2-1/q}(qx).
\label{eq:expqq}
\end{equation}
Note that Equation \ref{eq:2d-qgauss} is recognized as a
two-dimensional $q$-normal distribution,
\begin{equation}
p_{X,Y}(x,y)= \frac{1}{2\pi}\exp_{r}\Bigl(-\frac{1}{2+D(1-r)}(x^2+y^2)\Bigr), \\
\end{equation}
where $r=2-1/q$ and $D=2$. This is properly parameterized with each
marginal $q$-variance equal to one.
}

\textcolor{black}{
{\bf Proposition 2.}
The marginal distribution of $x$ is given by
\begin{eqnarray}
p_X(x) =
\left\{
\begin{array}{ll}
\frac{1}{B\Bigl(\frac{2-q'}{1-q'},\frac{1}{2}\Bigr)}\Bigl(\frac{1-q'}{3-q'}\Bigr)^{\frac{1}{2}}\Bigl[1-\frac{1-q'}{3-q'}x^2\Bigr]^{\frac{1}{1-q'}} & |x| \leq \sqrt{\frac{3-q'}{1-q'}} \quad (q' < 1)  \\
 \frac{1}{\sqrt{2\pi}}\exp\Bigl(-\frac{x^2}{2}\Bigr) & (q = q'=1)  \\
 \frac{1}{B\Bigl(\frac{1}{1-{q'}}-\frac{1}{2},\frac{1}{2}\Bigr)}\Bigl(\frac{{q'}-1}{3-{q'}}\Bigr)^{\frac{1}{2}}\Bigl[1+\frac{q'-1}{3-q'}x^2\Bigr]^{\frac{1}{1-q'}} & (1 < q' < 3) 
\end{array}
\right.
\label{eq:q-gauss}
\end{eqnarray}
where $q' = (3q-1)/(q+1)$. Hence, $x$ in Equation \ref{eq:GBMM} gives
a $q'$-Gaussian random variable. 
}

\textcolor{black}{
{\bf Proof of Proposition 2.}
Integrating Equation \ref{eq:2-q-gauss} in terms of $y$, 
we obtain Equation \ref{eq:q-gauss}. In the case of $q=1$, 
we obviously obtain
\begin{eqnarray}
\nonumber
p_X(x) &=& \int_{-\infty}^{\infty}p_{X,Y}(x,y)dy \\
\nonumber
&=&
\frac{1}{2\pi}\int_{-\infty}^{\infty}\exp\Bigl(-\frac{1}{2}(x^2+y^2)\Bigr)\mbox{d}y \\
&=& \frac{1}{\sqrt{2\pi}}\exp\Bigl(-\frac{x^2}{2}\Bigr).
\end{eqnarray}
}

\textcolor{black}{
In the case of $1 < q < 3$, we have
\begin{eqnarray}
\nonumber
p_X(x) &=& \int_{-\infty}^{\infty}p_{X,Y}(x,y)dy \\
\nonumber
&=&
\frac{1}{2\pi}\int_{-\infty}^{\infty}\exp_{2-1/q}\Bigl(-\frac{q}{2}(x^2+y^2)\Bigr)\mbox{d}y
\\
\nonumber
&=&
\frac{1}{2\pi}\int_{-\infty}^{\infty}\Bigl(1-\frac{1-q}{2}(x^2+y^2)\Bigr)^{\frac{q}{1-q}}\mbox{d}y \\
&=&
\nonumber
\frac{1}{\pi}\int_0^{\infty}\Bigl(1-\frac{1-q}{2}(x^2+y^2)\Bigr)^{\frac{q}{1-q}}\mbox{d}y
\\
\nonumber
&& \Bigl(\zeta = 1 + \frac{q-1}{2}x^2\Bigr) \\
\nonumber
&=& \frac{1}{\pi}\zeta^{\frac{q}{1-q}}\int_0^{\infty}
\Bigl(1+\frac{q-1}{2\zeta}y^2\Bigr)^{\frac{q}{1-q}} \mbox{d}y \\
\nonumber
&& \Bigl(t = \frac{1}{\frac{q-1}{2\zeta}y^2+1}\Bigr) \\
\nonumber
&=&
  \frac{1}{\pi\sqrt{2(q-1)}}\zeta^{\frac{q}{1-q}+\frac{1}{2}}\int_{0}^{1}t^{\frac{q}{q-1}-\frac{3}{2}}(1-t)^{-\frac{1}{2}}\mbox{d}t
  \\
&=&
  \frac{1}{\pi\sqrt{2(q-1)}}B\Bigl(\frac{q}{q-1}-\frac{1}{2},\frac{1}{2}\Bigr)\Bigl(1+\frac{q-1}{2}x^2\Bigr)^{-\frac{q+1}{2(q-1)}}.
\label{eq:q-gauss-derivation1}
\end{eqnarray}
Using the equality among beta function and gamma functions 
\begin{equation}
B(a,b)=\frac{\Gamma(a)\Gamma(b)}{\Gamma(a+b)},
\end{equation}
where the gamma function is defined as
\begin{equation}
\Gamma(a) = \int_0^{\infty}t^{a-1}e^{-t}\mbox{d}t,
\end{equation}
and $\pi = \Gamma(\frac{1}{2})^2$,
\begin{equation}
\nonumber
\frac{B\bigl(\frac{q}{q-1}-\frac{1}{2},\frac{1}{2}\bigr)}{\pi} = \frac{\Gamma(\frac{q}{q-1}-\frac{1}{2})\Gamma(\frac{1}{2})}{\Gamma(\frac{1}{2})^2\Gamma(\frac{q}{q-1})} = \frac{\Gamma(\frac{q}{q-1}-\frac{1}{2})}{\Gamma(\frac{1}{2})\Gamma(\frac{q}{q-1})}.
\end{equation}
Since we have $\Gamma(\frac{q}{q-1}) = (\frac{q}{q-1}-1)\Gamma(\frac{q}{q-1}-1) = \frac{1}{q-1}\Gamma(\frac{q}{q-1}-1)$, we get
\begin{equation}
\nonumber
\frac{B\bigl(\frac{q}{q-1}-\frac{1}{2},\frac{1}{2}\bigr)}{\pi} = 
\frac{(q-1)\Gamma(\frac{q}{q-1}-\frac{1}{2})}{\Gamma(\frac{1}{2})\Gamma(\frac{q}{q-1}-1)} = \frac{q-1}{B\bigl(\frac{q}{q-1}-1,\frac{1}{2}\bigr)}.
\end{equation}
Therefore, Equation \ref{eq:q-gauss-derivation1} can be rewritten as
\begin{equation}
p_X(x) = \frac{1}{B(\frac{q}{q-1}-1,\frac{1}{2})}\Bigl(\frac{q-1}{2}\Bigr)^{\frac{1}{2}}\Bigl(1+\frac{q-1}{2}x^2\Bigr)^{-\frac{q+1}{2(q-1)}}.
\end{equation}
Setting $q = \frac{q'+1}{3-q'}$, we obtain
\begin{equation}
p_X(x) =  \frac{1}{B\Bigl(\frac{1}{1-{q'}}-\frac{1}{2},\frac{1}{2}\Bigr)}\Bigl(\frac{{q'}-1}{3-{q'}}\Bigr)^{\frac{1}{2}}\Bigl[1+\frac{q'-1}{3-q'}x^2\Bigr]^{\frac{1}{1-q'}}
\end{equation}
}

\textcolor{black}{
In the case of $q < 1$, we obtain the joint density $p_{X,Y}(x,y)$ has
a compact support ranging from $-\sqrt{\frac{2}{1-q}x^2}$ to 
$\sqrt{\frac{2}{1-q}x^2}$.
\begin{eqnarray}
\nonumber
p_X(x) &=& \int_{-\sqrt{\frac{2}{1-q}x^2}}^{\sqrt{\frac{2}{1-q}x^2}}p_{X,Y}(x,y)dy \\
\nonumber
&=& \frac{1}{2\pi}\int_{-\sqrt{\frac{2}{1-q}x^2}}^{\sqrt{\frac{2}{1-q}x^2}}\Bigl(
1-\frac{1-q}{2}(x^2+y^2)\Bigr)^{\frac{q}{1-q}}\mbox{d}y \\
\nonumber
&=&\frac{1}{\pi}\int_0^{\sqrt{\frac{2}{1-q}x^2}}\Bigl(1-\frac{1-q}{2}(x^2+y^2)\Bigr)^{\frac{q}{1-q}}\mbox{d}y
\\
\nonumber
&& \Bigl(\zeta = 1-\frac{1-q}{2}x^2\Bigr) \\
\nonumber
&=&
\frac{1}{\pi\sqrt{2(q-1)}}\zeta^{\frac{q}{1-q}+\frac{1}{2}}\int_{1}^{\infty}t^{\frac{q}{q-1}-\frac{3}{2}}(1-t)^{-\frac{1}{2}}\mbox{d}t
\\
\nonumber
&& \Bigl(t = \frac{1}{s}\Bigr)
\\
\nonumber
&=& 
-\frac{1}{\pi\sqrt{2(q-1)}}\zeta^{\frac{q}{1-q}+\frac{1}{2}}\int_{1}^{0}s^{\frac{q}{1-q}}(1-s)^{-\frac{1}{2}}\mbox{d}s
\nonumber
\\
&=&
\frac{1}{\pi\sqrt{2(q-1)}}B\Bigl(\frac{1}{1-q},\frac{1}{2}\Bigr)\Bigl(1-\frac{1-q}{2}x^2\Bigr)^{\frac{2q+1}{2(1-q)}}.
\end{eqnarray}
Similarly to the case of $1 < q < 3$ setting $q = \frac{q'+1}{3-q'}$, we obtain
\begin{equation}
p_X(x) = \frac{1}{B\Bigl(\frac{2-q'}{1-q'},\frac{1}{2}\Bigr)}\Bigl(\frac{1-q'}{3-q'}\Bigr)^{\frac{1}{2}}\Bigl[1-\frac{1-q'}{3-q'}x^2\Bigr]^{\frac{1}{1-q'}},
\end{equation}
where $|x| \leq \sqrt{\frac{3-q'}{1-q'}}$.
}

\begin{figure}[!t]
\centering
\includegraphics[scale=0.6]{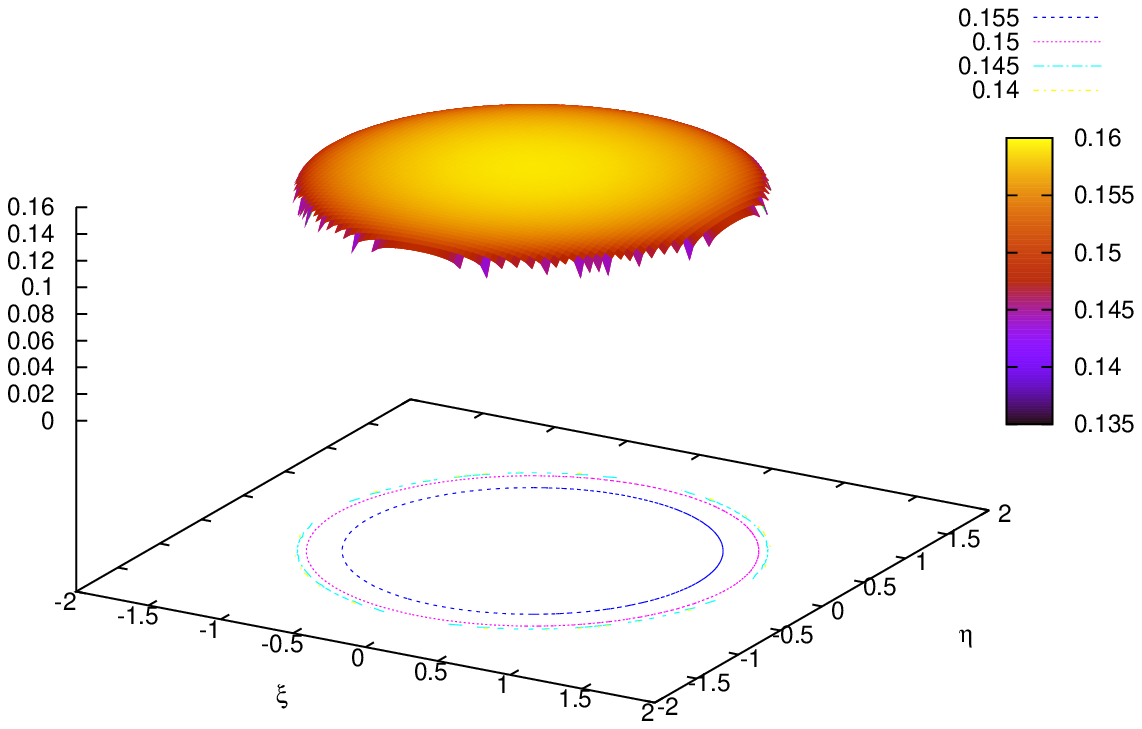}(a)
\includegraphics[scale=0.6]{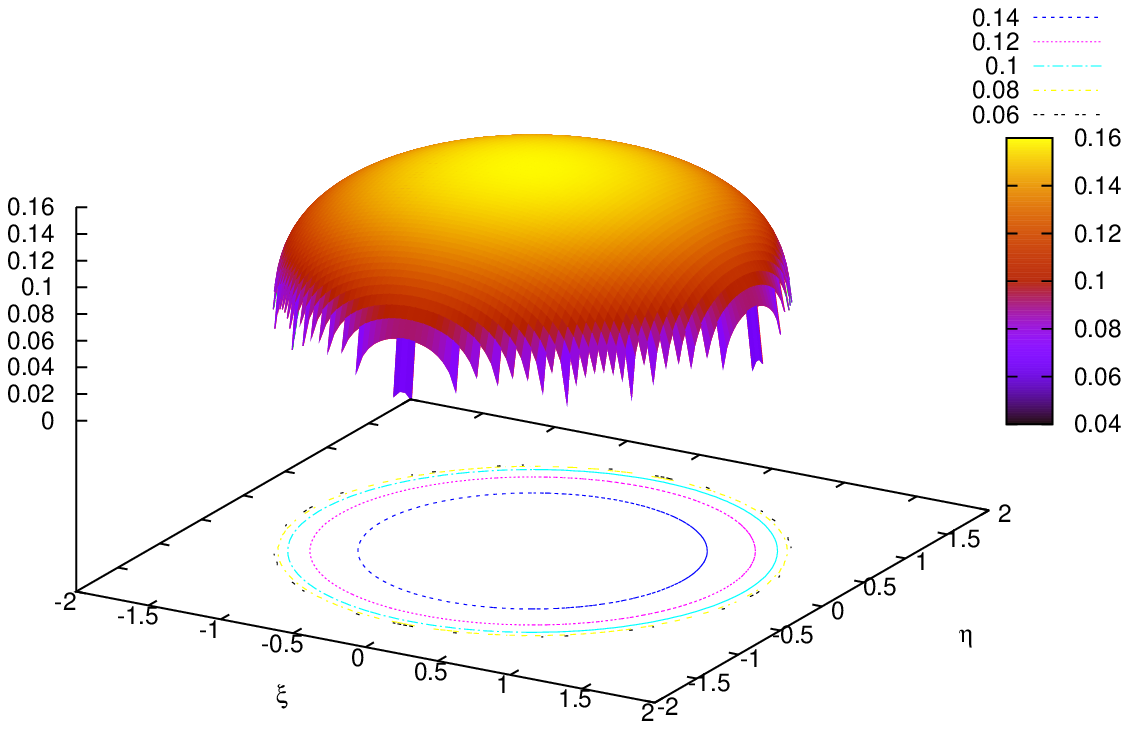}(b)
\includegraphics[scale=0.6]{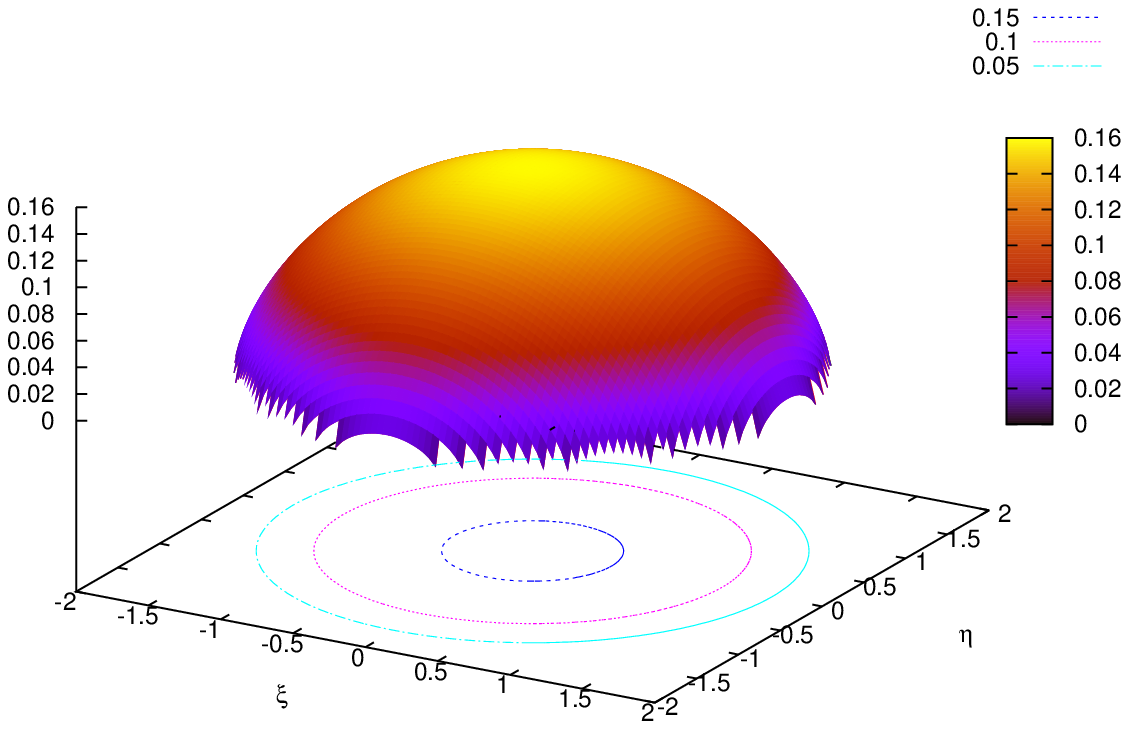}(c)
\includegraphics[scale=0.6]{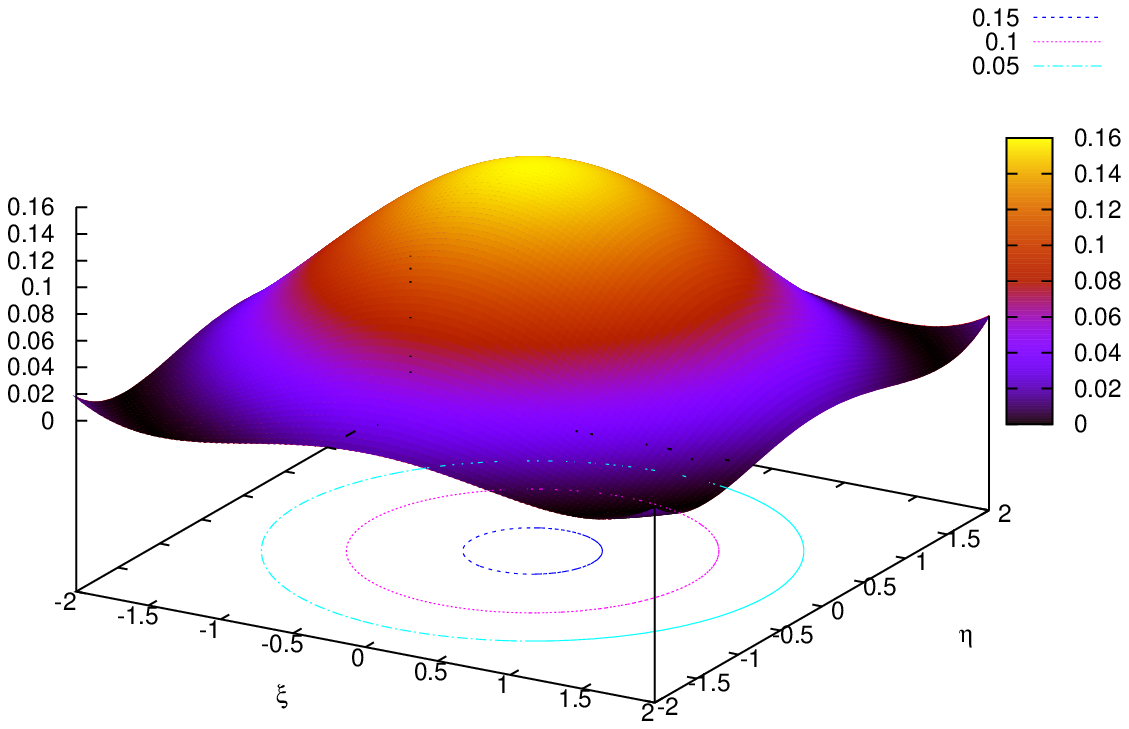}(d)
\includegraphics[scale=0.6]{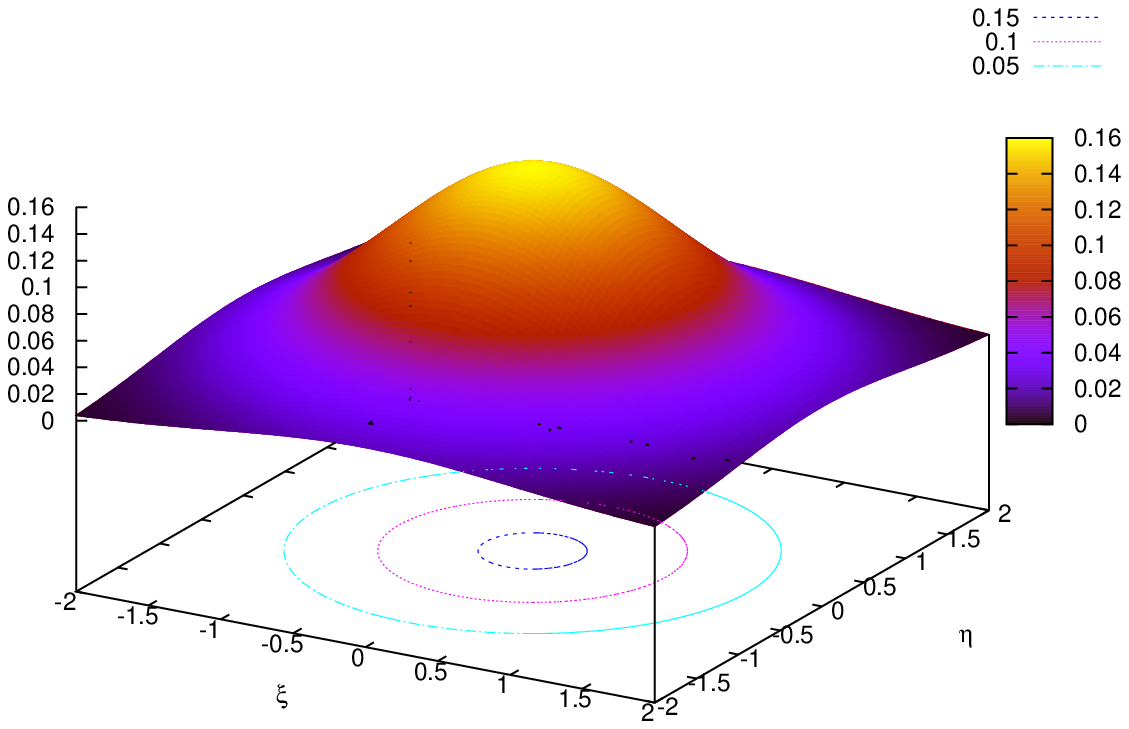}(e)
\includegraphics[scale=0.6]{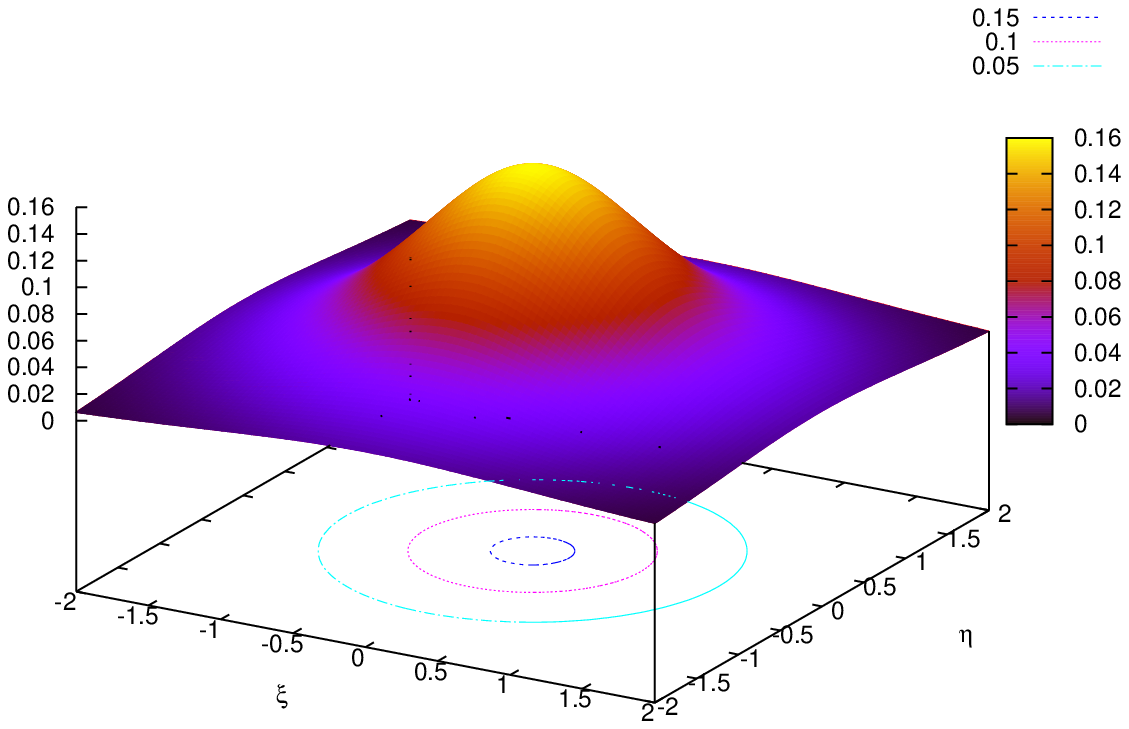}(f)
\includegraphics[scale=0.6]{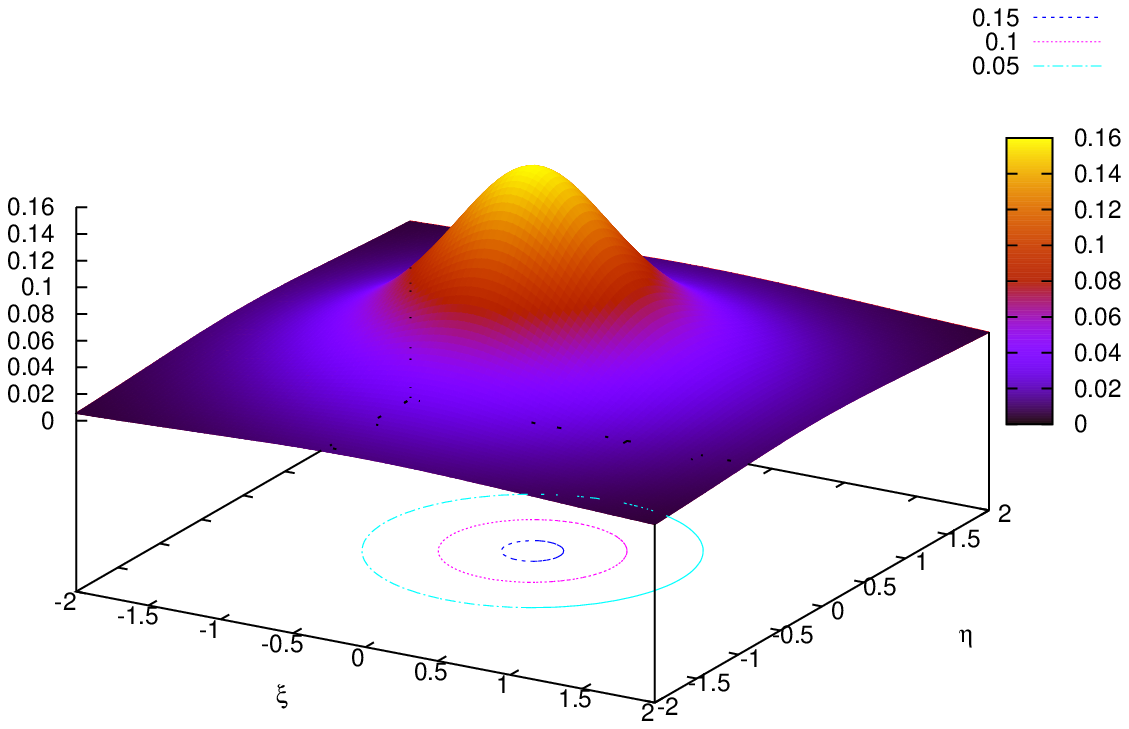}(g)
\caption{Three dimensional plots of joint density in terms of $\xi$ and
 $\eta$ for (a) $q'=-0.9$, (b) $-0.4$, (c) $0.1$, (d) $0.6$, (e) $1.1$
 ($\textcolor{black}{\nu}=19$), (f) $1.6$ ($\textcolor{black}{\nu}=2.33$), and 
(g) $2.1$ ($\textcolor{black}{\nu}=0.818$).}
\label{fig:3d-qgauss}
\end{figure}

\textcolor{black}{
Figure \ref{fig:3d-qgauss} shows the distribution of Equation
\ref{eq:2-q-gauss} for several cases of $q$. The distribution is the
spinning object. The marginal distribution in terms of $\xi$ is also
equivalent to Equation \ref{eq:q-gauss}. 
}

\section{Map dynamics}
Adler and Rivlin considered $P_d(w)=\cos d\theta$, where $w=\cos\theta$,
$0\leq \theta \leq \pi$, defined by Chebyshev polynomial of degree
$d$~\cite{Adler:64}, where $d$ is an integer. Clearly, $P_d(w)$ is
permutable $P_{d_1}(P_{d_2}(w)) = P_{d_1d_2}(w)$. They proved that for $d
\geq 2$ the ergodic invariant measure of the map dynamics $w_{n+1} = P_d(w_n)$
\textcolor{black}{has} \textcolor{black}{an explicit density function} 
\textcolor{black}{invariant} measure $\mu_W(w) =
\frac{1}{\pi\sqrt{1-w^2}}$. 

More generally, let us extend the Chebyshev polynomial to
a two-dimensional case as \cite{Umeno:06}.

\textcolor{black}{
{\bf Definition 4.}
$P_d(w)$ and $Q_d(w,v)$ are given as real and imaginary parts of
binomial expansion,   
\begin{equation}
(w+\textcolor{black}{i}v)^d = P_d(w) + \textcolor{black}{i} Q_d(w,v),
\end{equation}
where the equality $w^2+v^2=1$ is necessary in order to obtain $P_d(w)$
from this expansion. Here, in this definition we used the
\textcolor{black}{Eular}
equality
\begin{equation}
(\exp(\textcolor{black}{i}\theta))^d = \exp(\textcolor{black}{i}d\theta)
= \cos d\theta + \textcolor{black}{i}\sin d\theta.
\label{eq:Euler}
\end{equation} 
}

This $P_d(w)$ is the Chebyshev polynomial of degree
$d$. \textcolor{black}{The first few polynomials are explicitly 
given by $P_1(w)=w$, $Q_1(w,v)=v$, $P_2(w)=2w^2-1$,
$Q_2(w,v)=2wv$, $P_3(w)=4w^3 - 3w$, $Q_3(w,v)=v(4w^2-1)$,
$P_4(w)=8w^4 - 8w^2 + 1$, $Q_4(w,v)=v(8w^3-4w)$,
$P_5(w)=16 w^5 - 20 w^3 + 5w$, $Q_5(w,v) = v(16w^4-12w^2+1)$,
$P_6(w)=32w^6 - 48w^4 + 18w^2 - 1$, $Q_6(w,v)=v(32w^5-32w^3+6w)$, 
$P_7(w)=64 w^7 - 112 w^5 + 56 w^3 - 7w$,
$Q_7(w,v)=v(64w^6-80w^4+24w^2-1)$, $P_8(w) = 128w^8 - 256w^6 + 160w^4 -
32w^2 + 1$, and $Q_8(w,v)=v(128w^7-192w^5+80w^3-8w)$.}

\textcolor{black}{
{\bf Definition 5.} 
For $d \geq 2$, we define the map dynamics 
\begin{eqnarray}
w_{n+1}&=&P_d(w_n) 
\label{eq:real}
\\
v_{n+1}&=&Q_d(w_n,v_n)
\label{eq:imag}
\end{eqnarray}
on the unit circle $w_n^2+v_n^2=1$. \textcolor{black}{The set of
variables} $(w_{n},v_{n})$ is uniformly distributed on the unit circle
if we set an initial condition of $(w_0,v_0)$ on the unit circle. We
set \textcolor{black}{$v_0$} as an arbitrary value in $(0,1)$ and
\textcolor{black}{$w_0$} is given by \textcolor{black}{$w_0 = \pm\sqrt{1-v_0^2}$}.}

\textcolor{black}{
{\bf Lemma 1.}
The joint density of ergodic invariant measure for $w_{n+1} = P_d(w_n)$
and $v_{n+1} = Q_d(w_n,v_n)$ follows 
\begin{equation}
p_{W,V}(w,v) =
\frac{\delta\bigl(\sqrt{w^2+v^2}-1\bigr)}{2\pi\sqrt{w^2+v^2}}.
\label{eq:density-circle}
\end{equation}
}

\textcolor{black}{
{\bf Proof of Lemma 1.}
In addition to $P_d(w)=\cos d\theta$, we introduce $Q_d(w,v)=\sin
d\theta$, where $w=\cos\theta$ and $v=\sin\theta$, $0 \leq \theta\leq 2
\pi$. From the equality given in Equation \ref{eq:Euler}, the angular
$\theta_n$ of $w_n+iv_n$ follows the map dynamics
\begin{equation}
\theta_{n+1}=d\theta_n \quad \mbox{mod}\quad 2\pi,
\label{eq:theta-map}
\end{equation}
which is ergodic and has an ergodic density function~\cite{Arnol'd}
\begin{equation}
p_{\Theta}(\theta) = \frac{1}{2\pi} \quad (0 < \theta < 2\pi).
\label{eq:theta}
\end{equation}
Transforming the orthogonal coordinate $(w,v)$ into the polar
coordinate $(a,\theta)$ by $w = a\cos\theta$ and $v = a\sin\theta$, we
have $p_A(a) = \delta(a-1)$. Since ons has $a=\sqrt{w^2+v^2}$
$\frac{\partial w}{\partial a} = \cos\theta$, $\frac{\partial
  w}{\partial \theta} = -a\sin\theta$, $\frac{\partial v}{\partial a}
= \sin\theta$, and $\frac{\partial v}{\partial \theta} = a\cos\theta$,
the Jacobian matrix is expressed as
\begin{eqnarray}
\nonumber
\Bigl|\frac{\partial(\theta,a)}{\partial(w,v)}\Bigr| &=&
\left|
\begin{array}{cc}
\cos\theta & -a\sin\theta \\
\sin\theta & a\cos\theta 
\end{array}
\right|^{-1} \\
&=& a^{-1} = \frac{1}{\sqrt{w^2+v^2}}.
\end{eqnarray}
Therefore, the joint density of the ergodic invariant
measure of $w$ and $v$ can be described as
\begin{eqnarray}
\nonumber
p_{W,V}(w,v) &=&
p_{\Theta}(\theta)p_{A}(a)\Bigl|\frac{\partial(\theta,a)}{\partial(w,v)}\Bigr|
\\
&=& \frac{\delta(\sqrt{w^2+v^2}-1)}{2\pi\sqrt{w^2+v^2}}.
\label{eq:joint-wv}
\end{eqnarray}
}

\textcolor{black}{
{\bf Lemma 2.}
The density functions of ergodic invariant measure of Equation
\ref{eq:real}. and Equation \ref{eq:imag}, respectively, have the
form:
\begin{eqnarray}
\mu_W(w) &=& \frac{1}{\pi\sqrt{1-w^2}},
\label{eq:muw} \\
\mu_V(v) &=& \frac{1}{\pi\sqrt{1-v^2}}.
\label{eq:muv}
\end{eqnarray}
}

\textcolor{black}{
{\bf Proof of Lemma 2.}
From Equation \ref{eq:joint-wv} we can calculate $p_{W}(w)$ and
$p_{V}(v)$ as the marginal distribution in terms of $w$ and
$v$. Integrating Equation \ref{eq:joint-wv} with respect to $v$ and
$w$, we respectively obtain
\begin{eqnarray}
\nonumber
p_W(w) &=& \int_{-\infty}^{\infty}p_{WV}(w,v)\mbox{d}v \\
\nonumber
       &=&
\int_{-\infty}^{\infty}\frac{\delta\bigl(\sqrt{w^2+v^2}-1\bigr)}{2\pi\sqrt{w^2+v^2}} 
\mbox{d}v \\
       &=& \frac{1}{\pi\sqrt{1-w^2}}, \\
\nonumber
p_V(v) &=& \int_{-\infty}^{\infty}p_{WV}(w,v)\mbox{d}w \\
\nonumber
       &=&
\int_{-\infty}^{\infty}\frac{\delta\bigl(\sqrt{w^2+v^2}-1\bigr)}{2\pi\sqrt{w^2+v^2}}
\mbox{d}w \\
       &=& \frac{1}{\pi\sqrt{1-v^2}}.
\end{eqnarray}
}

\textcolor{black}{
{\bf Definition 6.} 
As an alternative method for generating $q$-Gaussian random variables,
we propose chaotic maps based on the following map dynamics:
\begin{equation}
\left\{
\begin{array}{lll}
w_{n+1} &=& \textcolor{black}{P_d(w_n)} \\
v_{n+1} &=& \textcolor{black}{Q_d(w_n, v_n)} \\
z_{n+1} &=& f_{l,c}(z_n)
\end{array}
\right.,
\label{eq:circle}
\end{equation}
where
\begin{eqnarray}
f_{l,c}(z) &=& g \circ \textcolor{black}{\underbrace{T_l \circ \cdots \circ T_l }_c} \circ g^{-1}(z),
\label{eq:z-map} \\
w_n^2 + v_n^2 &=& 1, \qquad z_n > 0
\end{eqnarray}
assuming 
\begin{eqnarray}
g(u) &=& \sqrt{-2\ln_q(u)}, 
\label{eq:trans-lnq}
\\
g^{-1}(z) &=& \exp_q\Bigl(-\frac{z^2}{2}\Bigr),
\label{eq:trans-expq}
\end{eqnarray}
where $T_l(u)$ is an $l$-th order piecewise linear map defined as
\begin{equation}
T_l(u) = 
\left\{
\begin{array}{lc}
\left\{
\begin{array}{ll}
lx & (0 \leq u < 1/l) \\
-lx + 2 & (1/l \leq < 2/l) \\
\vdots & \\
lx - (l-1) & (1-1/l \leq x < 1) \\
\end{array}
\right. & (l: \mbox{odd}) \\
\left\{
\begin{array}{ll}
lx & (0 \leq u < 1/l) \\
-lx + 2 & (1/l \leq < 2/l) \\
\vdots & \\
-lx + l & (1-1/l \leq x < 1) \\
\end{array}
\right. & (l: \mbox{even})
\end{array}
\right. 
\label{eq:picewise-map}
\end{equation}
For example, in the case of $l=2$, Equation
\ref{eq:picewise-map} gives 
the tent map
\begin{eqnarray}
\nonumber
T_2(u) &=&
\left\{
\begin{array}{ll}
2u & (0 \leq u < 1/2) \\
-2u+2 & ( 1/2 \leq u < 1)
\end{array}
\right.\\
&=& 1 - |1-2u|.
\label{eq:tent-map}
\end{eqnarray}
In the case of $l=3$, Equation \ref{eq:picewise-map} is expressed as
\begin{equation}
T_3(u) =
\left\{
\begin{array}{ll}
3u & (0 \leq u < 1/2) \\
-3u+2 & (1/2 \leq u < 1) \\
3u-2 & (2/3 \leq u < 1)
\end{array}
\right.
\end{equation}
The number of iteration $c$ is an integer greater than or equal to
1. The order $l$ of the piecewise linear map is an integer greater
than or equal to 2. By using the product among $z_n$, $w_n$, and $v_n$, 
\begin{equation}
\left\{
\begin{array}{lll}
\xi_{n} &=& z_n w_n \\
\eta_{n} &=& z_n v_n \\
\end{array}
\right.,
\label{eq:map}
\end{equation}
we can also obtain two-dimensional deterministic dynamics. The random
seed of this pseudo random generator is given by $(v_0, z_0)$, where
we set $w_0$ as $w_0=\pm\sqrt{1-v_0^2}$.}

\textcolor{black}{
Note that factor 2 in front of $q$-exponential function in Equation
\textcolor{black}{\ref{eq:tent-map}} should be replaced with a value
both smaller than and close to 2, such as 1.99999, for the round
error correction in the case of actual numerical computation.
}

\textcolor{black}{
{\bf Lemma 3.}
The density of ergodic invariant measure of $z_{n+1} = f_{l,c}(z_n)$
follows the one-side distribution, 
\begin{equation}
p_Z(z) = z\exp_{2-1/q}\Bigl(-\frac{q}{2}z^2\Bigr) \quad (z \geq 0).
\label{eq:density-z}
\end{equation}
}

\textcolor{black}{
{\bf Proof of Lemma 3.}
The density of the ergodic invariant measure~\cite{Arnol'd} of the 
piecewise linear map 
\begin{equation}
u_{n+1}=\textcolor{black}{\underbrace{T_l\circ \cdots \circ T_l}_c}(u_n),
\end{equation}
follows the uniform distribution $p_U(u)=1 \quad (0<u<1)$ independently of
$l$ and $c$. Since we obtain $\mbox{d}u/\mbox{d}z =
u^q\sqrt{-2\ln_q(u)}$ from the transformation in Equation
\ref{eq:trans-lnq}, we have
\begin{eqnarray}
\nonumber
p_Z(z) &=& p_U(z)\Bigl|\frac{\mbox{d}u}{\mbox{d}z}\Bigr| \\
\nonumber
&=& z\Bigl(\exp_q(-z^2/2)\Bigr)^q \\  
&=& z\exp_{2-1/q}\Bigl(-\frac{q}{2}z^2\Bigr) \quad (z \geq 0).
\end{eqnarray}
In this derivation, we used the equality introduced in Equation \ref{eq:expqq}.
}

\textcolor{black}{
{\bf Theorem 1.}
\textcolor{black}{The} joint density $p_{\Xi, H}(\xi, \eta)$ of ergodic
invariant measure of map dynamics Equation \ref{eq:map} is the
$q$-Gaussian distribution which is the same as
Equation \ref{eq:2-q-gauss} and given by
\begin{equation}
p_{\Xi,H}(\xi,\eta) = \frac{1}{2\pi}\exp_{2-1/q}\Bigl(-\frac{q}{2}(\xi^2+\eta^2)\Bigr).
\end{equation}
}

\textcolor{black}{
{\bf Proof of Theorem 1.}
By using Equation \ref{eq:density-circle} and Equation \ref{eq:density-z},
the joint density $p_{\Xi,H}(\xi,\eta)$ of the ergodic invariant measure
in terms of $\xi$ and $\eta$ is given as
\begin{eqnarray}
\nonumber
p_{\Xi,H}(\xi,\eta) &=&
\int_{0}^{\infty}p_{Z,W,V}(z,\xi/z,\eta/z)\Bigl|\frac{\partial
  (z,w,v)}{\partial (z,\xi,\eta)}\Bigr|\mbox{d}z \\
\nonumber
&=&
\int_{0}^{\infty}p_Z(z)p_{W,V}(\xi/z,\eta/z)z^{-2}dz \\
\nonumber
&=& \int_{0}^{\infty}
z^{-1}\exp_{2-1/q}\Bigl(-\frac{q}{2}z^2\Bigr) 
\frac{\delta\Bigl(\frac{1}{z}\sqrt{\xi^2+\eta^2}-1\Bigr)}{2\pi\sqrt{(\xi/z)^2+(\eta/z)^2}}\mbox{d}z \\
&=& 
\frac{1}{2\pi}\exp_{2-1/q}\Bigl(-\frac{q}{2}(\xi^2+\eta^2)\Bigr) 
\end{eqnarray}
}

\textcolor{black}{
{\bf Theorem 2.} The marginal density of $\xi$ is
a one-dimensional $q$-Gaussian distribution, 
\begin{eqnarray}
p_{\Xi}(\xi) =
\left\{
\begin{array}{ll}
\frac{1}{B\Bigl(\frac{2-q'}{1-q'},\frac{1}{2}\Bigr)}\Bigl(\frac{1-q'}{3-q'}\Bigr)^{\frac{1}{2}}\Bigl[1-\frac{1-q'}{3-q'}\xi^2\Bigr]^{\frac{1}{1-q'}} & |\xi| \leq \sqrt{\frac{3-q'}{1-q'}} \quad (q' < 1)  \\
 \frac{1}{\sqrt{2\pi}}\exp\Bigl(-\frac{\xi^2}{2}\Bigr) & (q = q'=1)  \\
 \frac{1}{B\Bigl(\frac{1}{1-{q'}}-\frac{1}{2},\frac{1}{2}\Bigr)}\Bigl(\frac{{q'}-1}{3-{q'}}\Bigr)^{\frac{1}{2}}\Bigl[1+\frac{q'-1}{3-q'}\xi^2\Bigr]^{\frac{1}{1-q'}} & (1 < q' < 3) 
\end{array}
\right.,
\end{eqnarray}
where $q' = (3q-1)/(q+1)$. Hence, sequences $\xi_n$ generated from the
maps in Definition 1. are random numbers sampled from a
$q'$-Gaussian distribution, where $q'=(3q-1)/(q+1)$.} 

\textcolor{black}{
{\bf Proof of Theorem. 2} From Proposition 2., the marginal
distribution of $\xi$ is the same functional form as Equation
\ref{eq:q-gauss}.}

\section{Numerical simulation}

\begin{figure}[!t]
\centering
\includegraphics[scale=0.6]{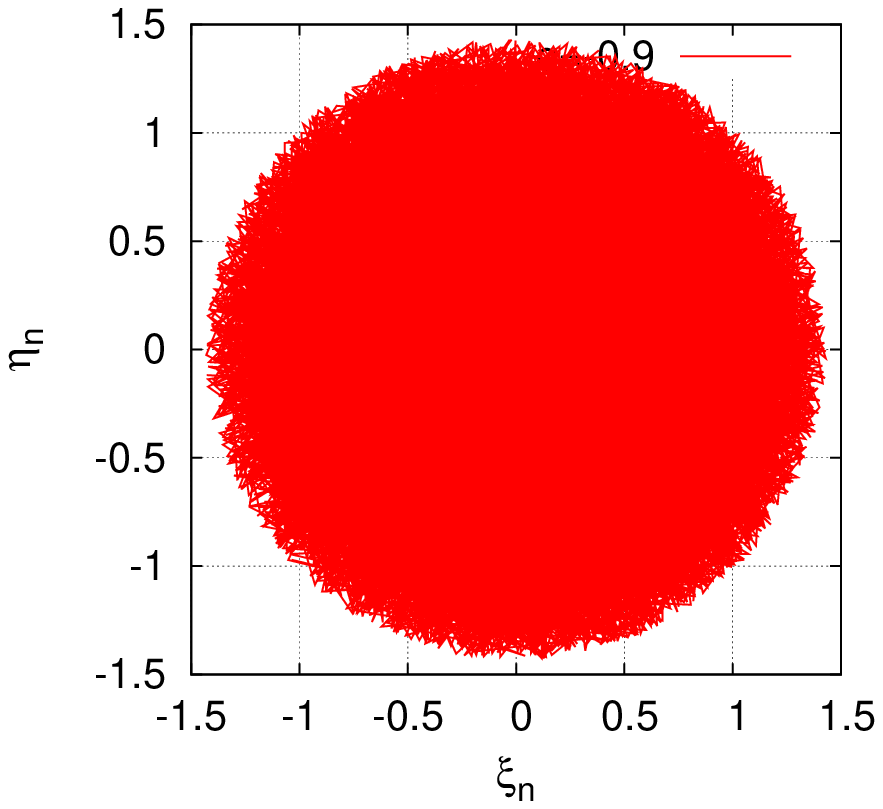}(a)
\includegraphics[scale=0.6]{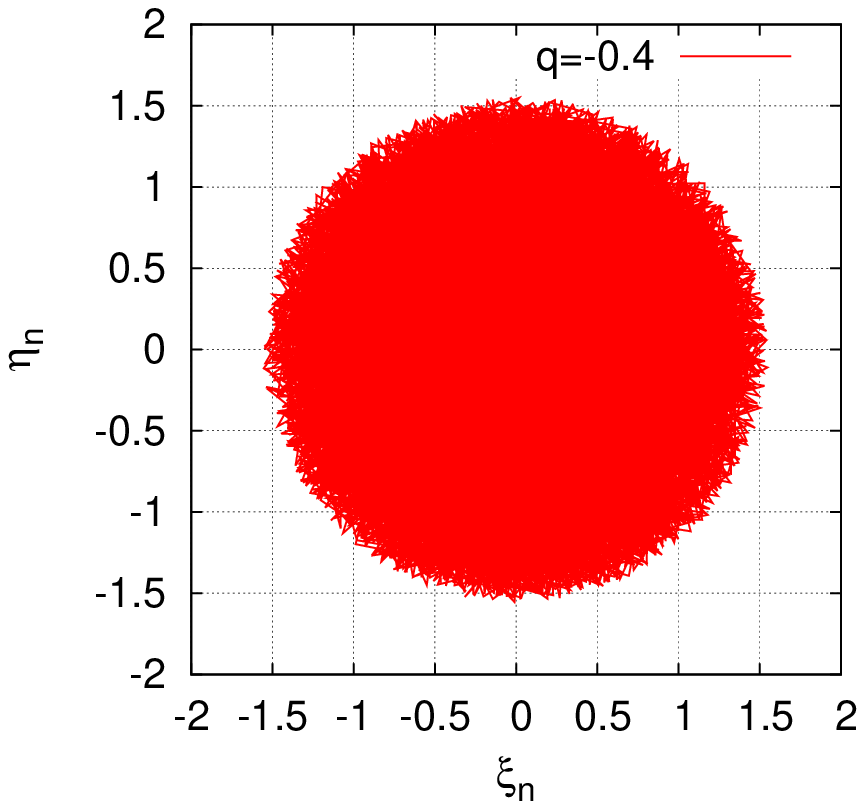}(b)
\includegraphics[scale=0.6]{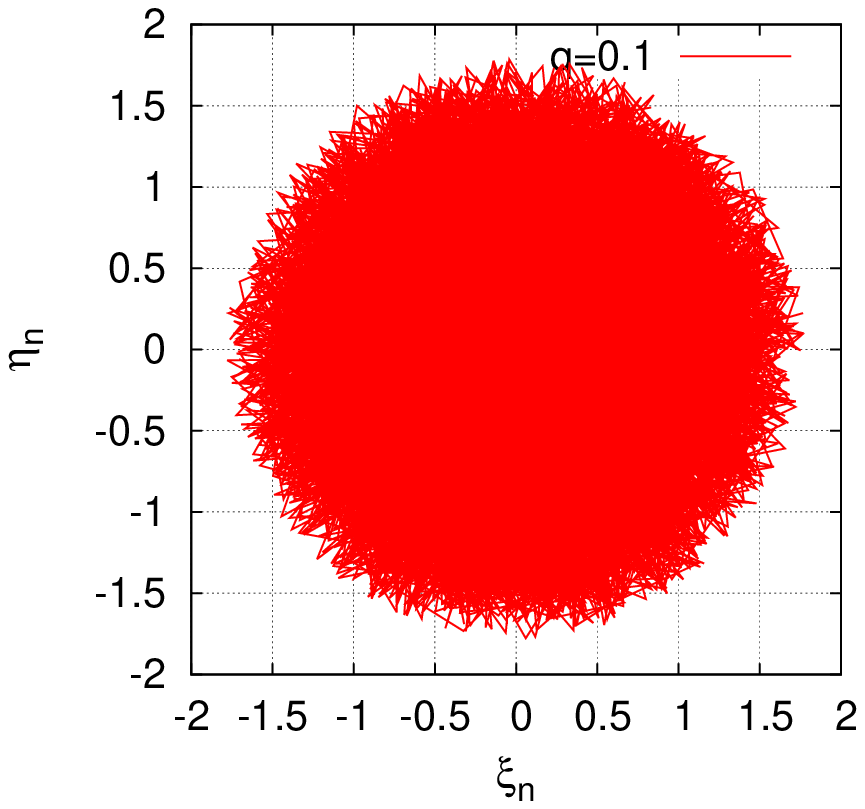}(c)
\includegraphics[scale=0.6]{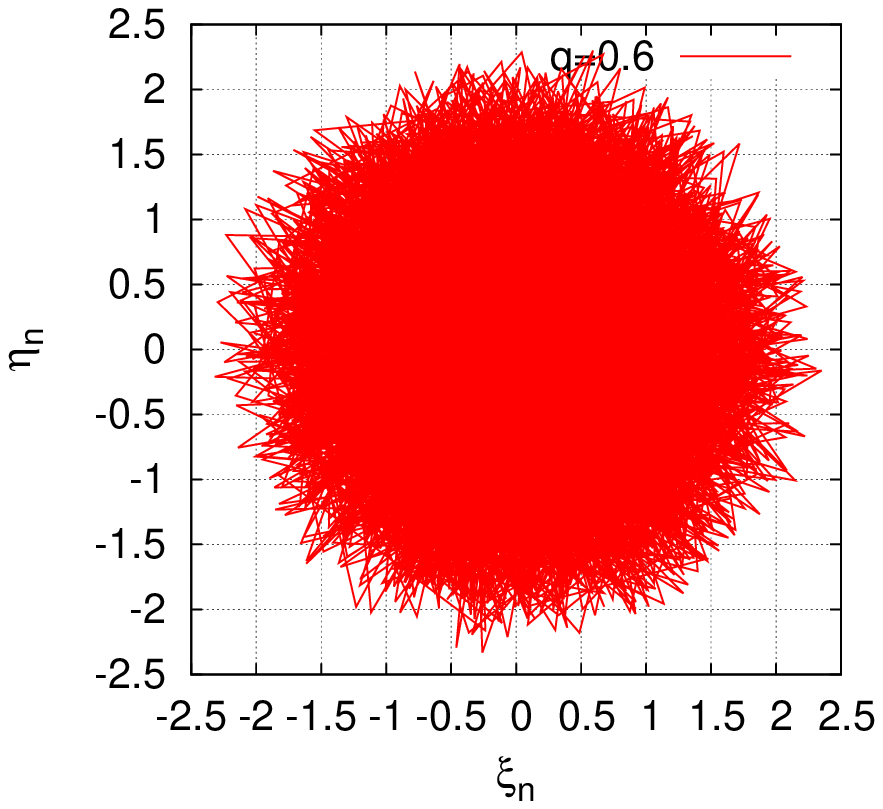}(d)
\includegraphics[scale=0.6]{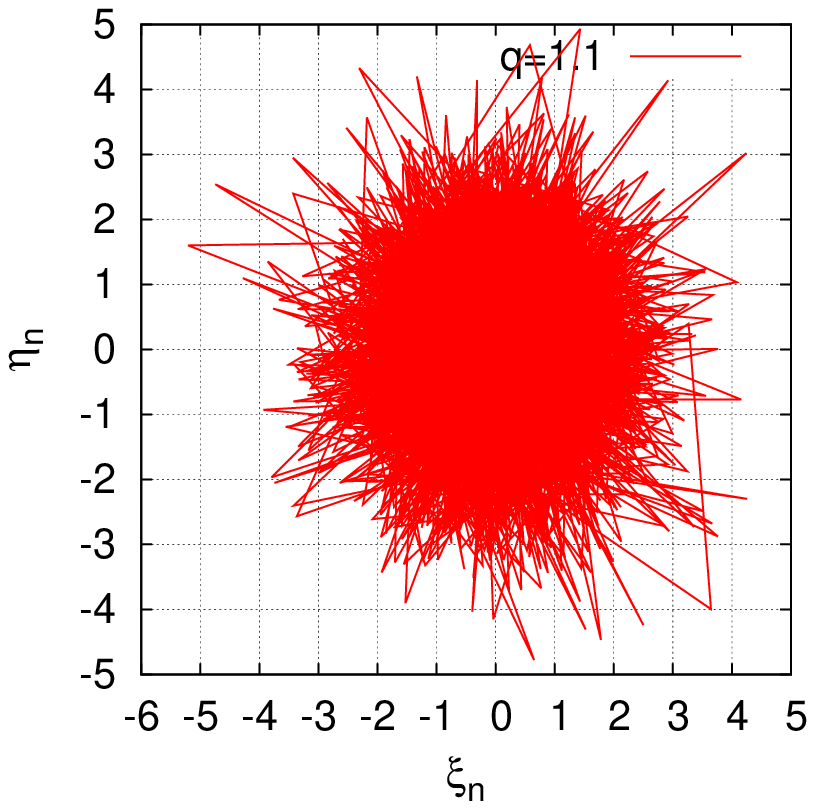}(e)
\includegraphics[scale=0.6]{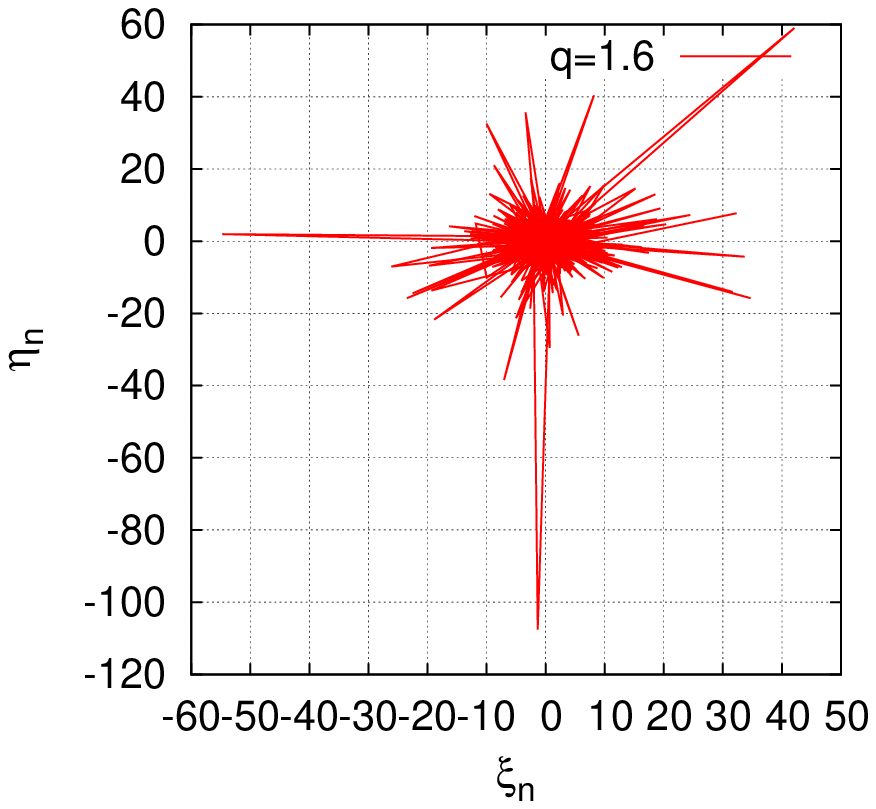}(f)
\includegraphics[scale=0.6]{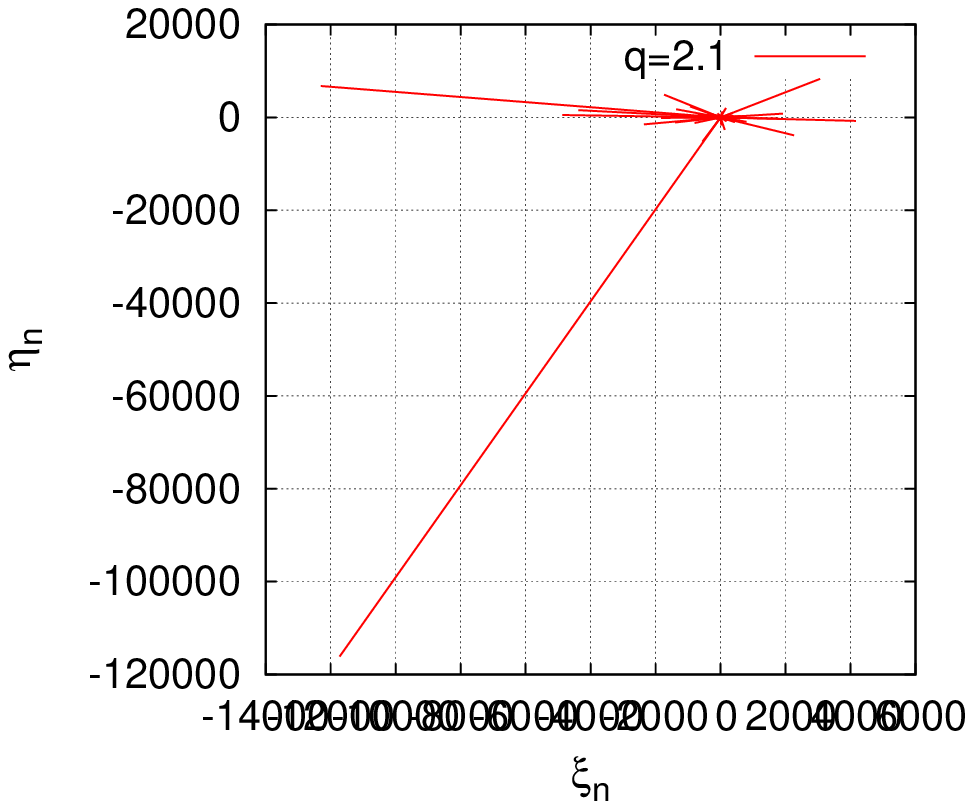}(g)
\caption{A sample path of the map dynamics at $d=8$, $l=2$, and $c=1$ for 
(a) $q'=-0.9$, (b) $-0.4$, (c) $0.1$, (d) $0.6$, (e) $1.1$
 ($\textcolor{black}{\nu}=19$), (f) $1.6$ ($\textcolor{black}{\nu}=2.33$),
 and (g) $2.1$ ($\textcolor{black}{\nu}=0.818$).}
\label{fig:seq}
\end{figure}

\begin{figure}[!t]
\centering
\includegraphics[scale=0.6]{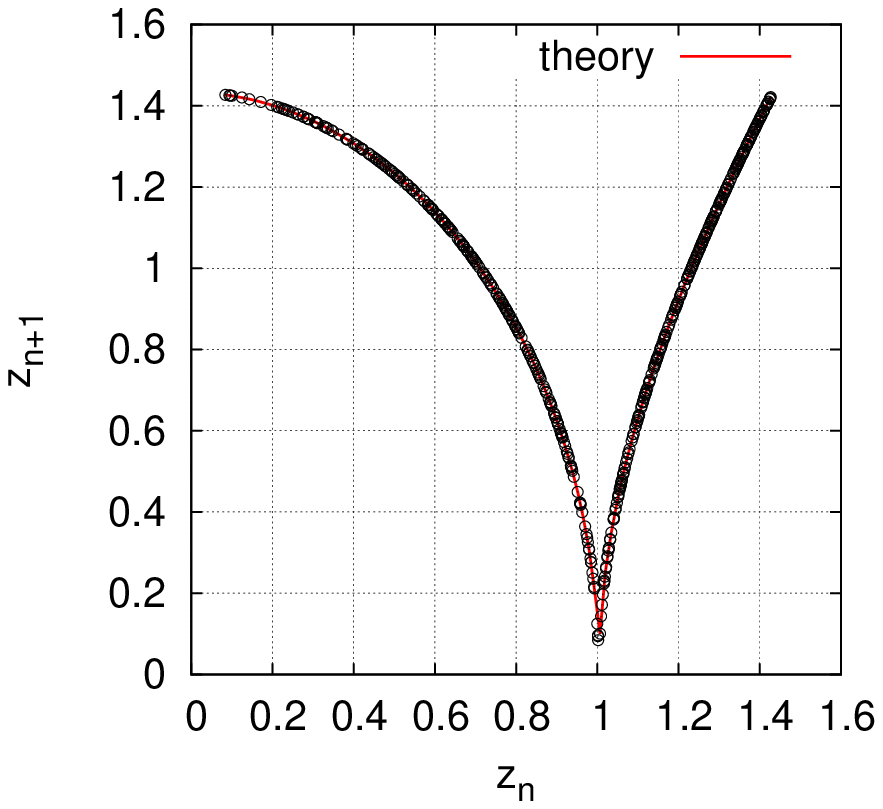}(a)
\includegraphics[scale=0.6]{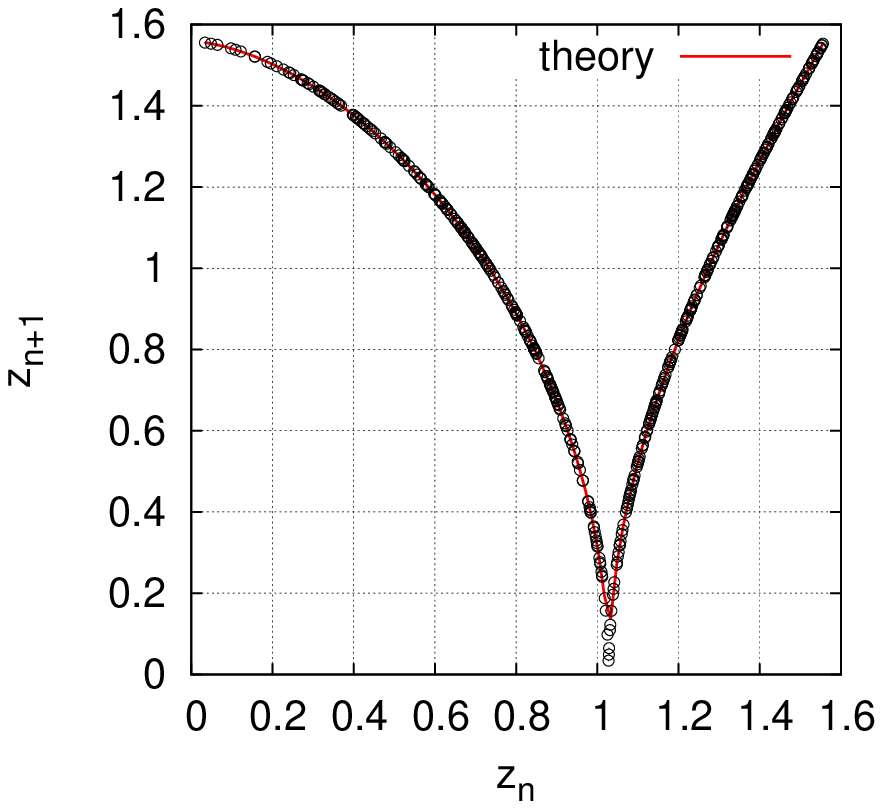}(b)
\includegraphics[scale=0.6]{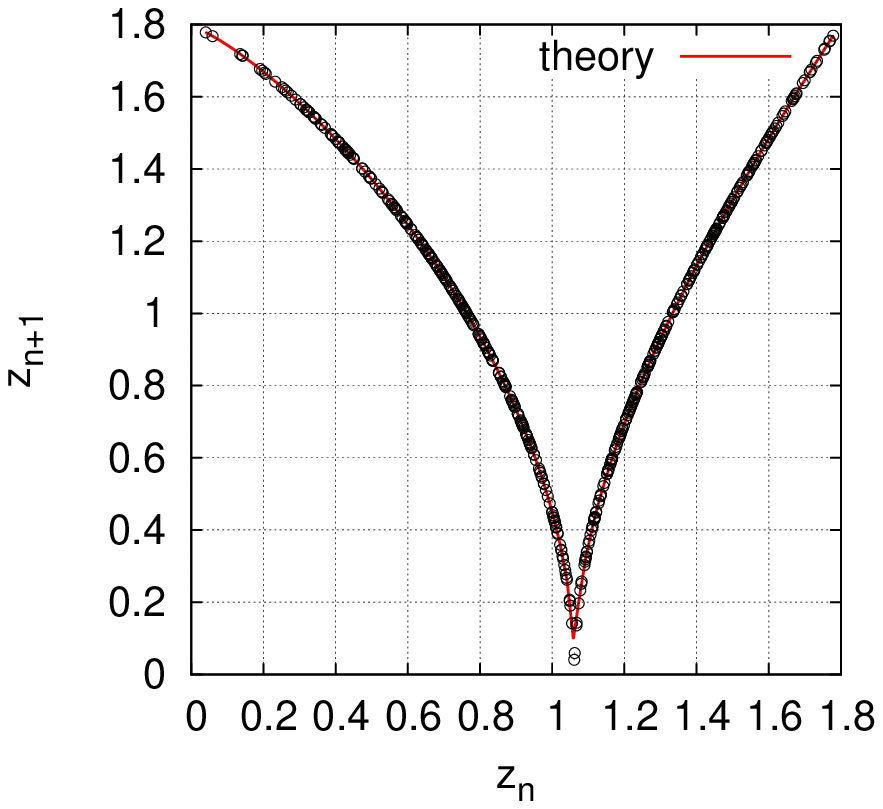}(c)
\includegraphics[scale=0.6]{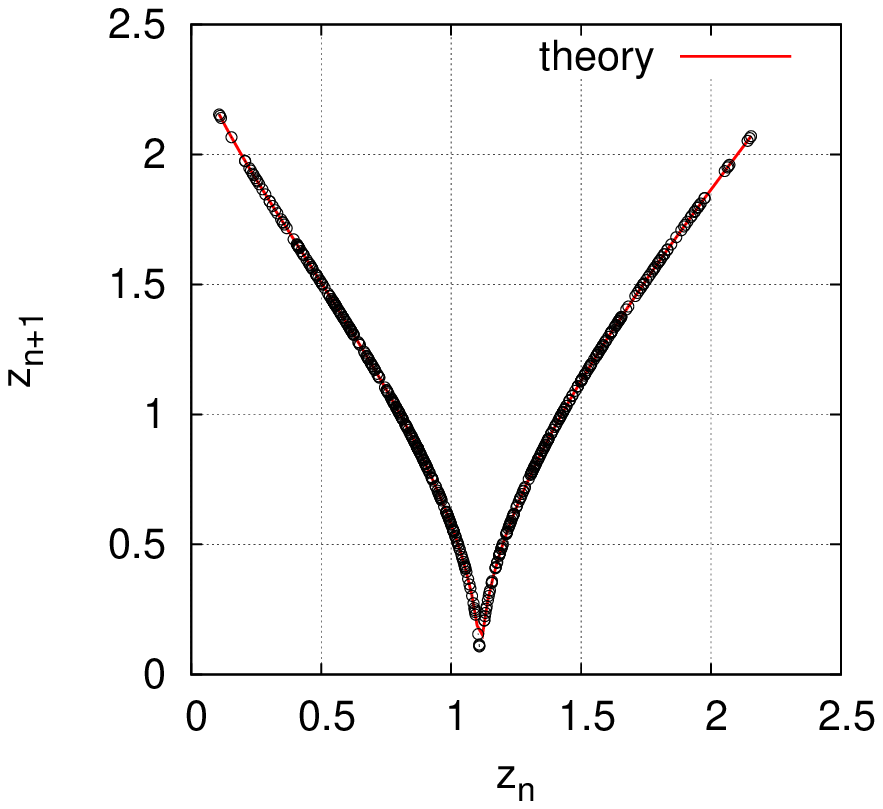}(d)
\includegraphics[scale=0.6]{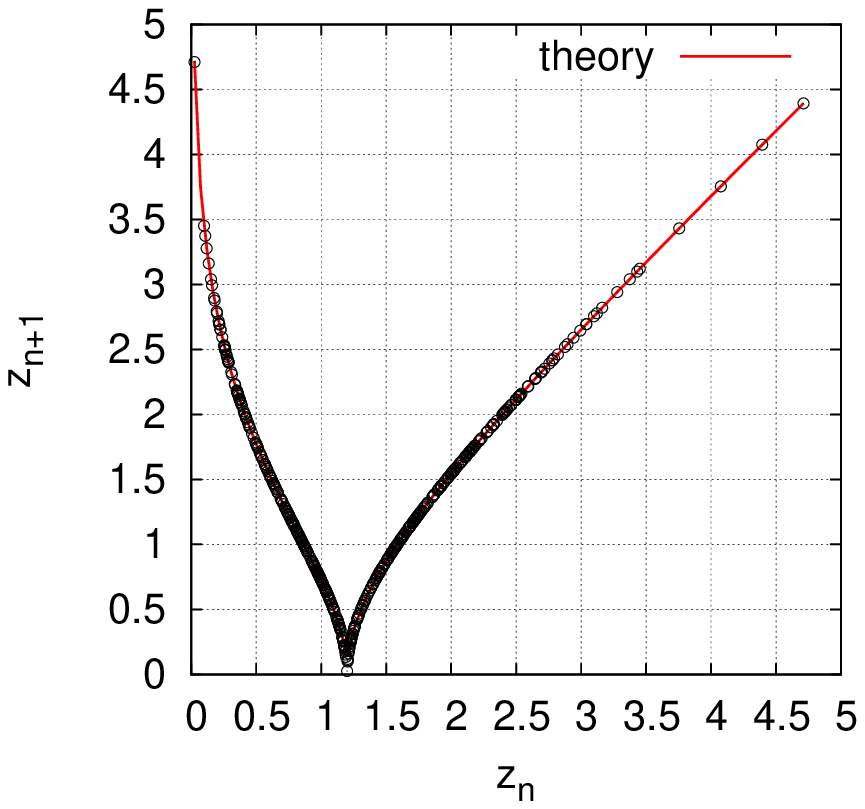}(e)
\includegraphics[scale=0.6]{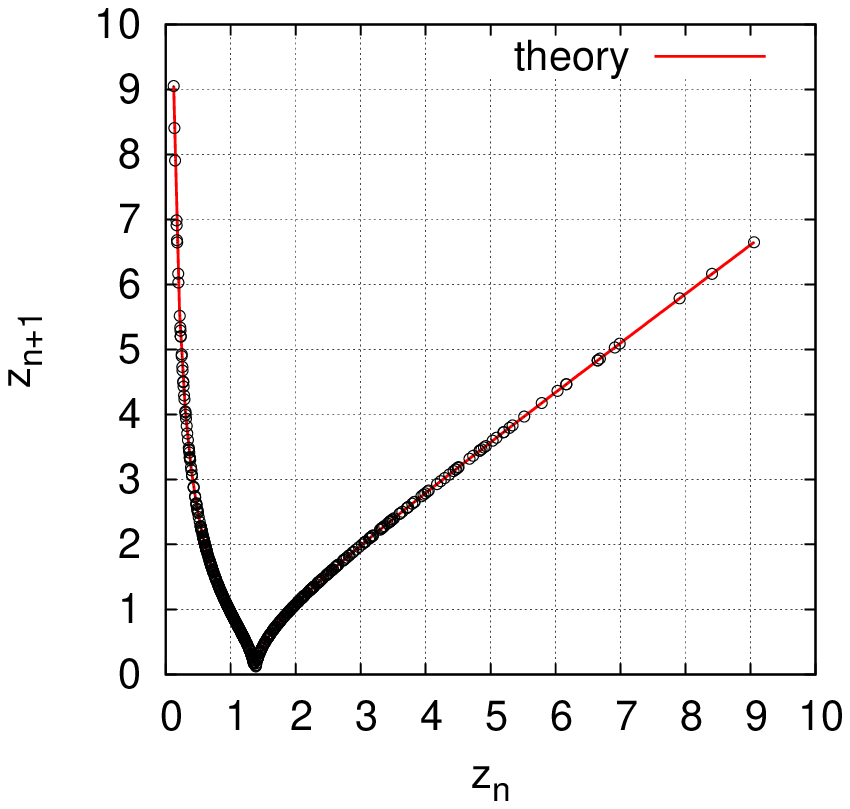}(f)
\includegraphics[scale=0.6]{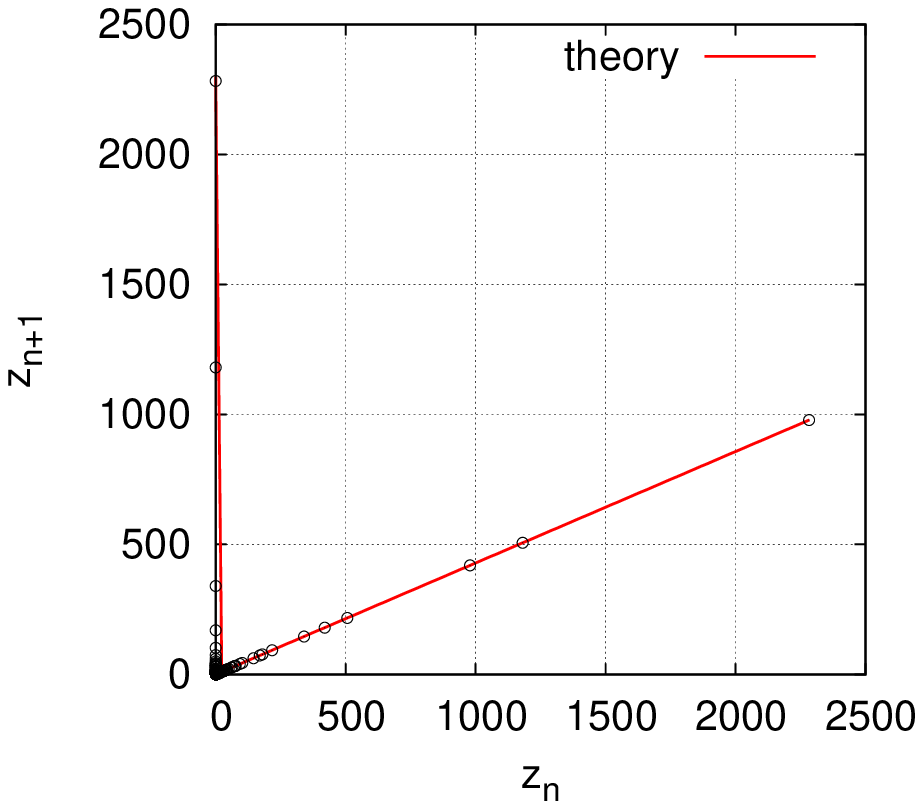}(g)
\caption{Return map between $z_{n}$ and $z_{n+1}$ for 
(a) $q'=-0.9$, (b) $-0.4$, (c) $0.1$, (d) $0.6$, (e) $1.1$
 ($\textcolor{black}{\nu}=19$), (f) $1.6$ ($\textcolor{black}{\nu}=2.33$),
  and (g) $2.1$ ($\textcolor{black}{\nu}=0.818$). The solid curve represents
  $z_{n+1}=f_{2,1}(z_n)$ for each value of $q=(q'+1)/(3-q')$.} 
\label{fig:map}
\end{figure}

Figure \ref{fig:seq} shows sample paths for several values of $q'$.
As shown in these figures, they seem to be from a trapped random walk to
L\'evy walk as $q$ is increasing.
Figure \ref{fig:map} shows the return maps between $z_{n+1}$ and $z_{n}$. 
They show the determinism of the \textcolor{black}{proposed} random 
\textcolor{black}{number} generator. The return map of $z_n$
\textcolor{black}{at $l=2$ and $c=1$} shows the functional form of the map
function introduced in
Equation \ref{eq:z-map}. \textcolor{black}{$f_{2,1}(z)=0$} holds at $z = 
\sqrt{-2\ln_q(1/2)}$. Since one has 
\begin{equation}
\frac{\mbox{d}f_{\textcolor{black}{2,1}}}{\mbox{d}z} =
\left\{
\begin{array}{l}
-\frac{2^{1-q}\bigl(1-\exp_q(-\frac{z^2}{2})\bigr)^{-q}\bigl(1+(1-q)(-\frac{z^2}{2})\bigr)^{\frac{q}{1-q}}z}{\sqrt{-2\ln_q[2(1-\exp_q(-\frac{z^2}{2}))]}}
 \\ \qquad \qquad \qquad (z < \sqrt{-2\ln_q(1/2)}) \\
\frac{2^{1-q}z}{\sqrt{-2\ln_q[2\exp_q(-\frac{z^2}{2})]}} \\ 
\qquad \qquad \qquad (z > \sqrt{-2\ln_q(1/2)}) \\
\end{array}
\right.,
\end{equation}
the Lyapunov exponent of $z_n$, defined as
\begin{equation}
\lambda = \lim_{t\rightarrow\infty}\frac{1}{t}\sum_{n=0}^{t-1}
 \log\Bigl|f'(z_n)\Bigr| = \log 2 = h_{KS}, 
\end{equation}
is computable. Here, $h_{KS}$ is \textcolor{black}{the} Kolmogrov-Sinai
entropy. The relation $\lambda = h_{KS}$ holds in one dimensional case
by the Pesin identity. Independently of the initial conditions $(v_0,
z_0)$ and the parameter $q$, it is numerically confirmed that the
Lyapunov exponent 
$\lambda$ approaches $\log(2)$ at $l=2$ and $c=1$. This is consistent
with the theoretical value \textcolor{black}{of chaotic map}, which is 
conjugate with a diffeomorphism $g$ for the tent
map. \textcolor{black}{More generally, the Lyapunov exponent $\lambda$
  approaches to $c\log(l)$ in a general case of $f_{l,c}$.} This
iterated map is deterministic, however, the auto-correlation function
of the productive variable $\xi = w z$,
\begin{equation}
C(m) = \lim_{t \rightarrow
 \infty}\frac{1}{t}\sum_{n=0}^{t-1}\xi_n\xi_{n+m}- \Bigl(\lim_{t \rightarrow
 \infty}\frac{1}{t}\sum_{n=0}^{t-1}\xi_n\Bigr)^2,
\label{eq:corr-xi}
\end{equation}
decays 0 for $m \geq 1$ from the orthogonality of the Chebyshev
polynomials. Obviously, the expectation value of $\xi$ is 
\begin{equation}
\lim_{t\rightarrow \infty}\frac{1}{t} \sum_{n=0}^{t-1}\xi_n
= \int_{-\infty}^{\infty}\xi p_{\Xi}(\xi) \mbox{d}\xi = 0.
\end{equation}

Since due to the independence of $w$ and $z$, we have
\begin{eqnarray}
\nonumber 
C(m) &=& \lim_{t\rightarrow
 \infty}\frac{1}{t}\sum_{n=0}^{t-1} w_n z_n \underbrace{P_d\circ\cdots \circ P_d}_{m}(w_n) \underbrace{f_{l,c}\circ\cdots\circ f_{l,c}}_m(z_n)\\
\nonumber
&=& \Bigl(\int_{-1}^{1}w 
\underbrace{P_d\circ\cdots \circ P_d}_{m}(w)
\mu_W(w)\mbox{d}w\Bigr)  \\
&\times& \Bigl(\int_{0}^{\infty}z 
\underbrace{f_{l,c}\circ\cdots\circ f_{l,c}}_m(z) p_z(z)\mbox{d}z\Bigr),
\end{eqnarray}
we obtain the auto-correlation of $\xi$ as
\begin{equation}
C(m) = 
\left\{
\begin{array}{ll}
\frac{1}{2}\delta_{1,2^m} B\Bigl(2,\frac{1}{1-q}\Bigr) & (q < 1) \\
\frac{1}{2}\delta_{1,2^m} & (q = 1) \\
\frac{1}{2}\delta_{1,2^m} B\Bigl(2,\textcolor{black}{\frac{3-q}{5-3q}}\Bigr) & (1 < q < \frac{5}{3}) 
\end{array}
\right..
\end{equation}
\textcolor{black}{Note that $C(0)$ is not finite for $5/3 < q < 3$ since
the variance of $q$-Gaussian distribution is not finite for $5/3 < q <
2$ and it is undefined for $2 < q < 3$.} In this derivation, we used the
permutability and the orthogonality of the Chebyshev polynomials,
\begin{eqnarray}
\nonumber
&&\int_{-1}^{1}\underbrace{P_{\textcolor{black}{j}}\circ\cdots\circ P_{\textcolor{black}{j}}}_n(w) 
\underbrace{P_{\textcolor{black}{k}}\circ\cdots\circ P_{\textcolor{black}{k}}}_m(w) \mu_W(w)\mbox{d}w  \\
\nonumber
&=&
\int_{-1}^{1}P_{\textcolor{black}{j}^n}(w) P_{\textcolor{black}{k}^m}(w) \mu_W(w)\mbox{d}w 
\\
&=&\frac{1}{\pi}\int_{0}^{\pi} \cos(\textcolor{black}{j}^n\theta)  \cos(\textcolor{black}{k}^m\theta) \mbox{d}\theta = \frac{1}{2}\delta_{\textcolor{black}{j}^{n},\textcolor{black}{k}^{m}}.
\end{eqnarray}
In the same way, it can be proved that the auto-correlation
function of the productive variable $\eta = v z$ also decays 0 for $m \geq 1$.

\begin{figure}[!t]
\centering
\includegraphics[scale=0.6]{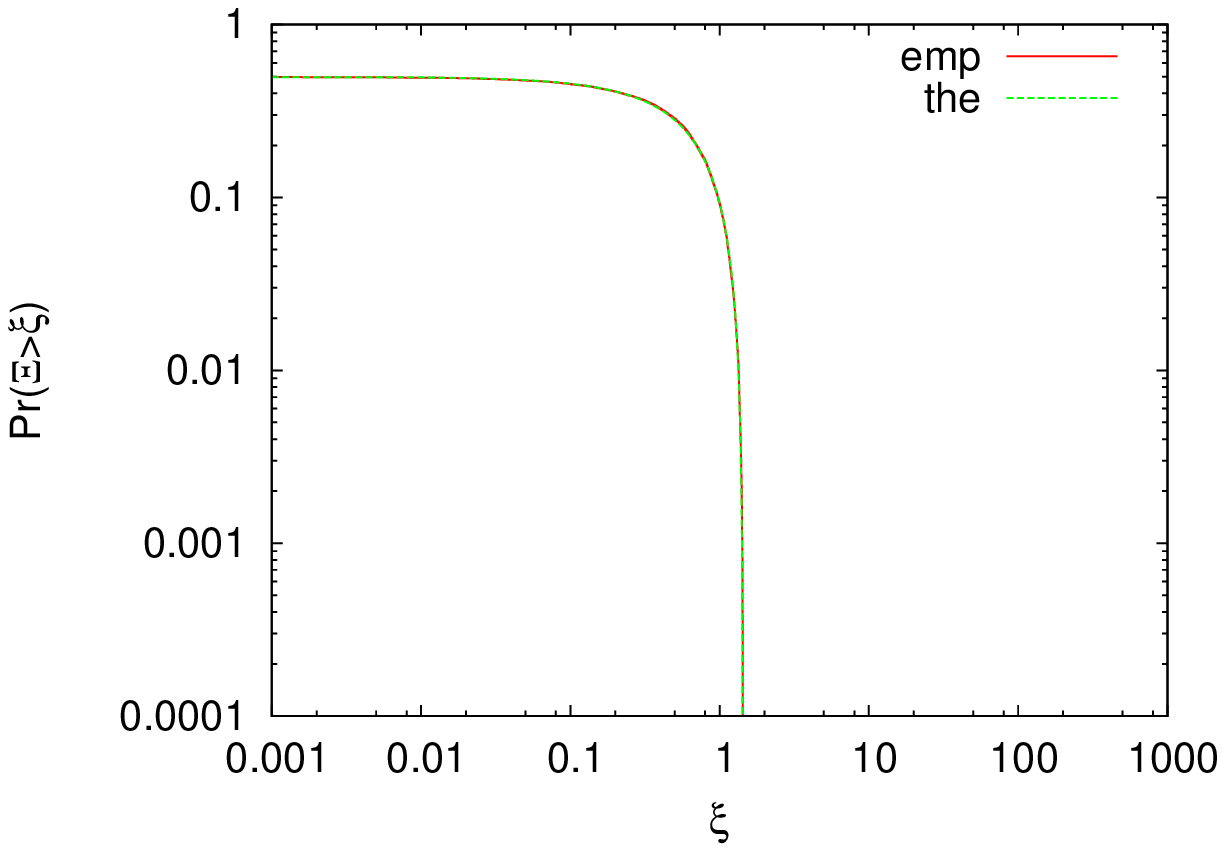}(a)
\includegraphics[scale=0.6]{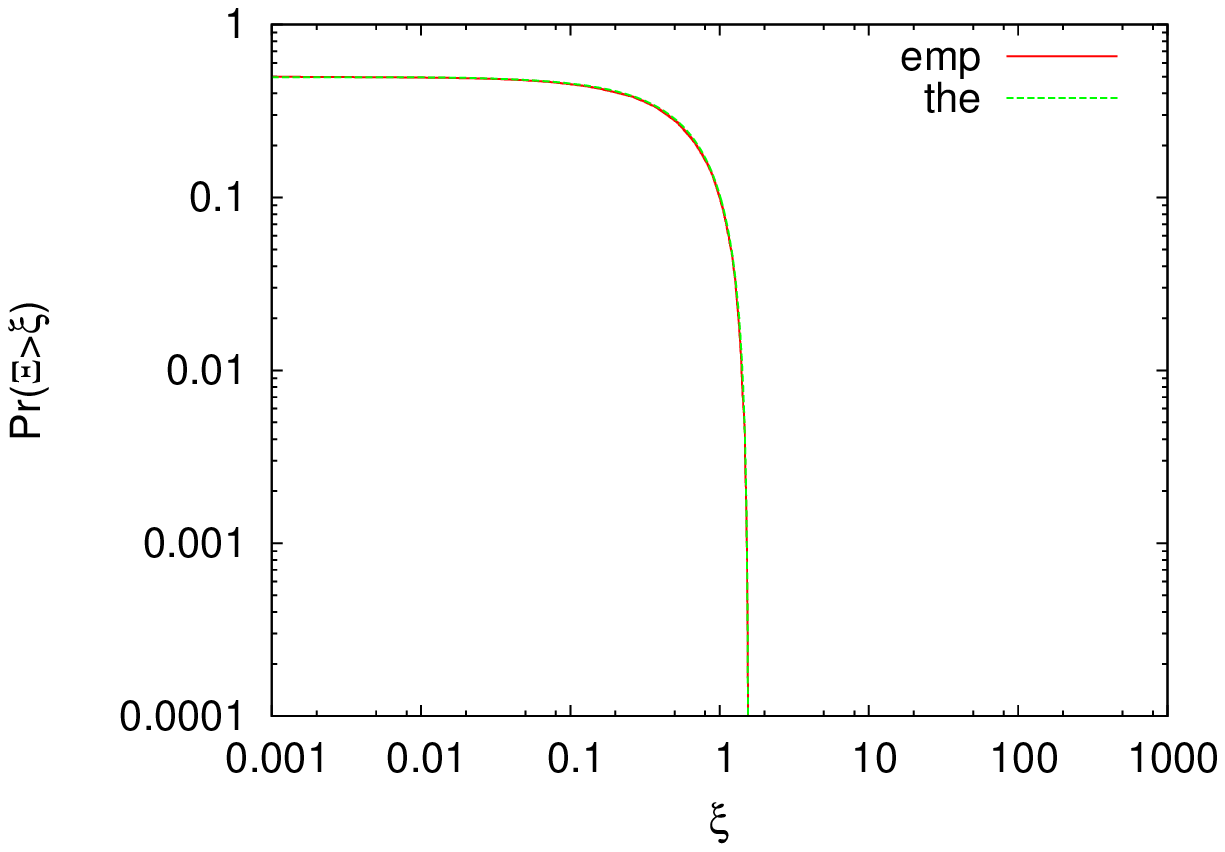}(b)
\includegraphics[scale=0.6]{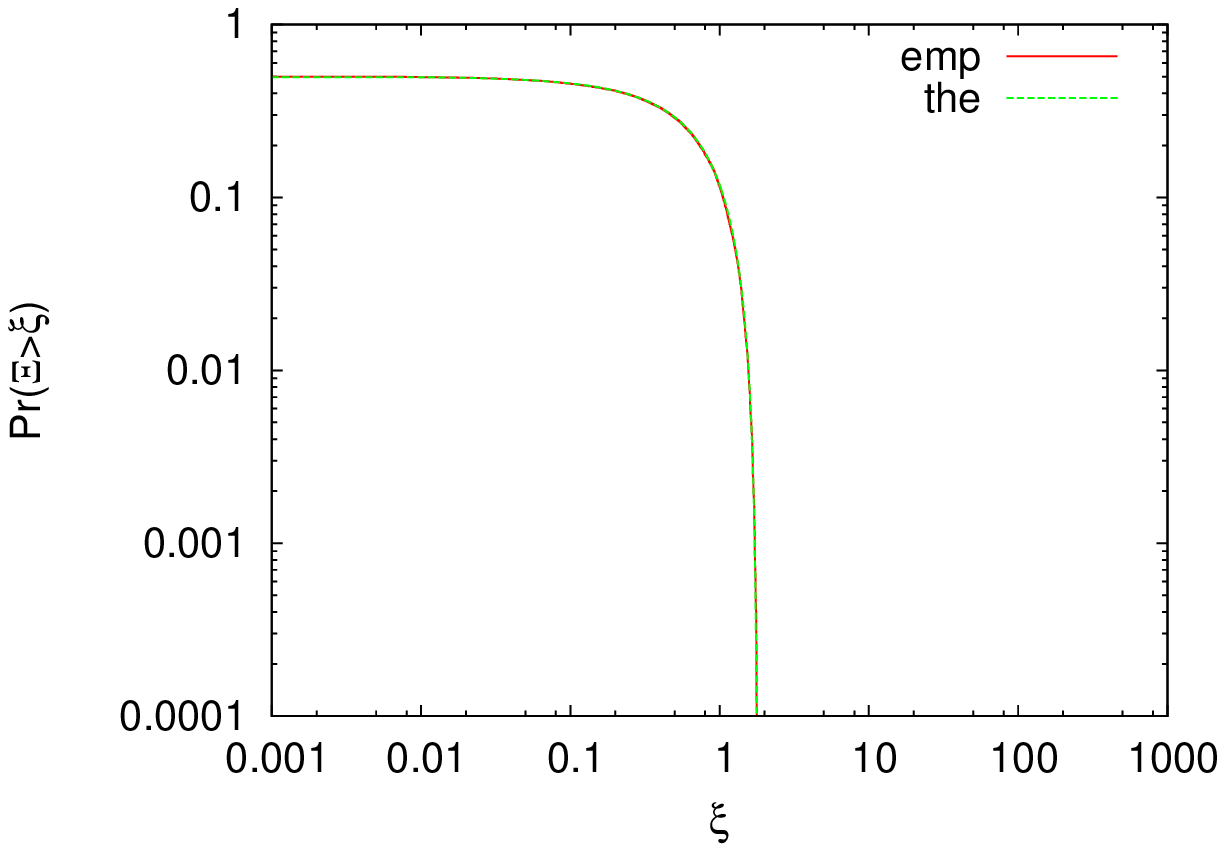}(c)
\includegraphics[scale=0.6]{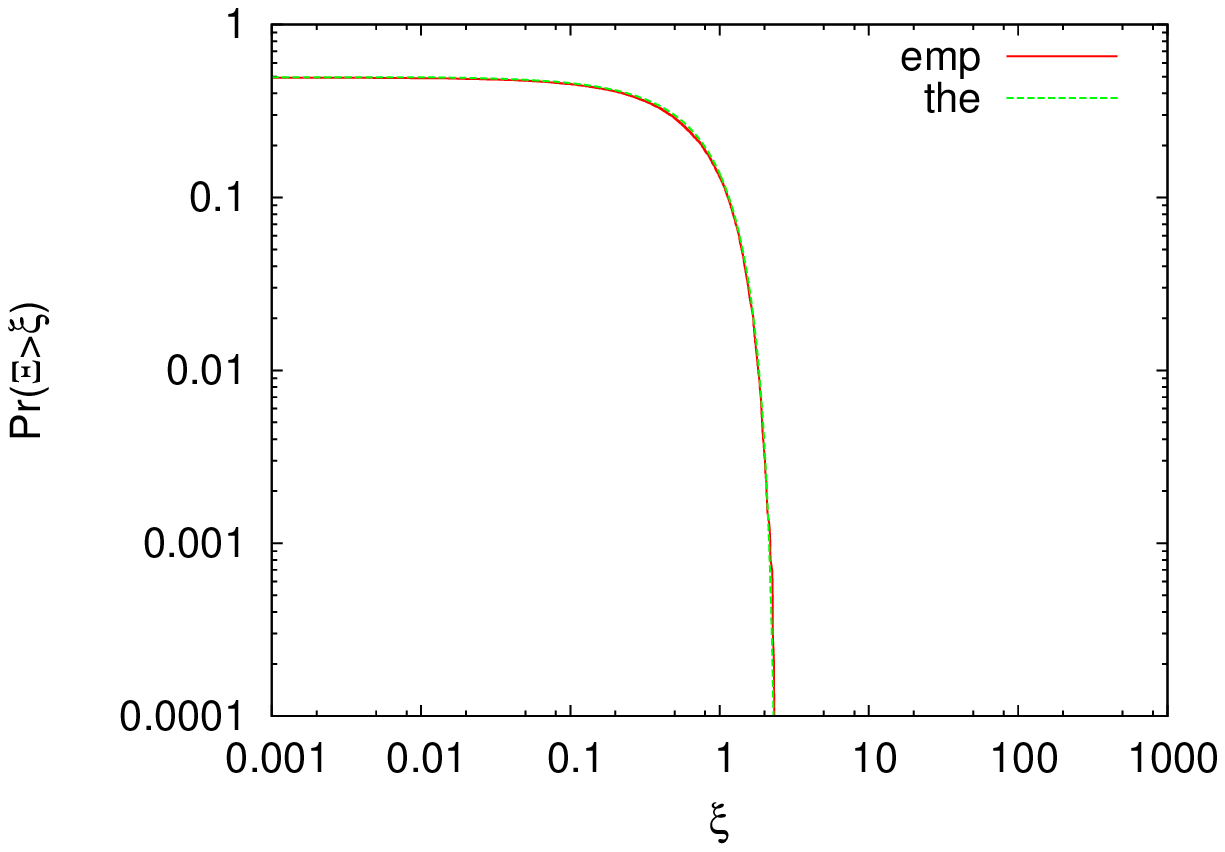}(d)
\includegraphics[scale=0.6]{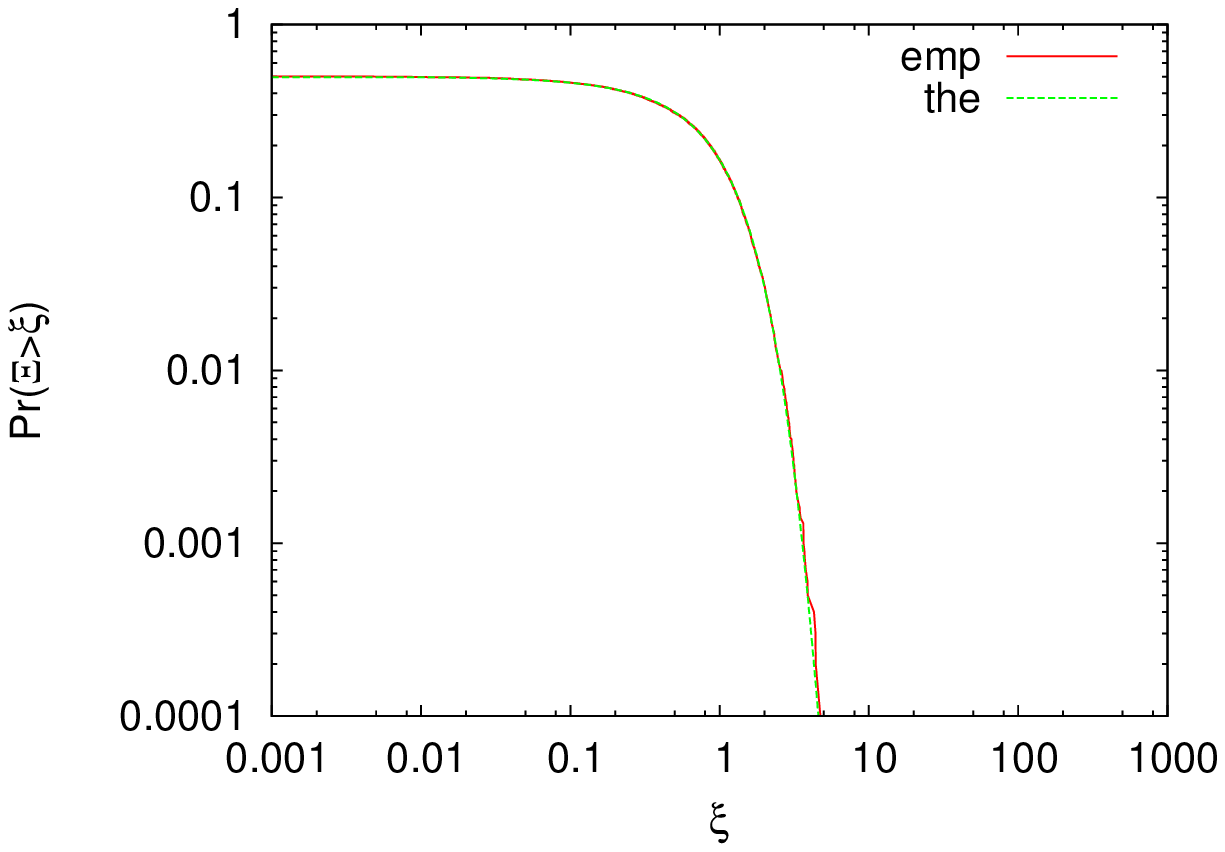}(e)
\includegraphics[scale=0.6]{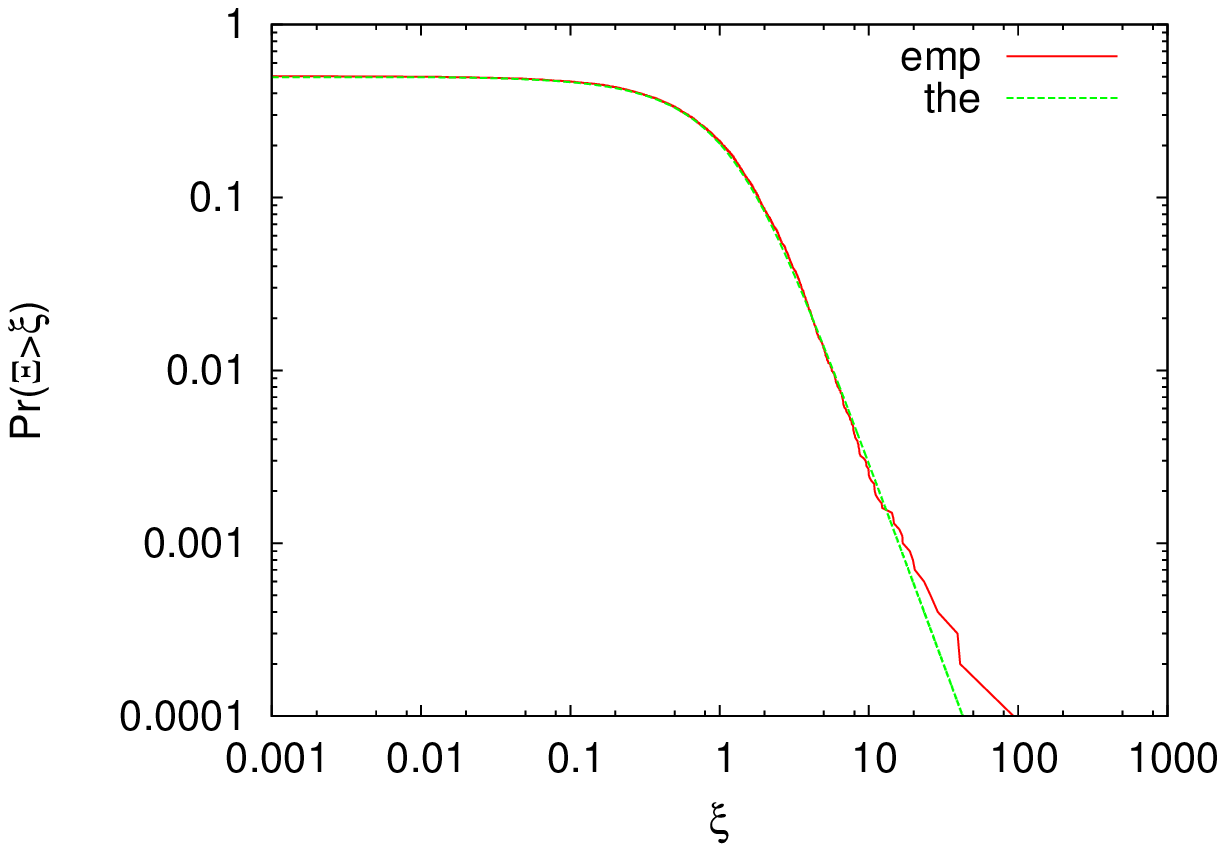}(f)
\includegraphics[scale=0.6]{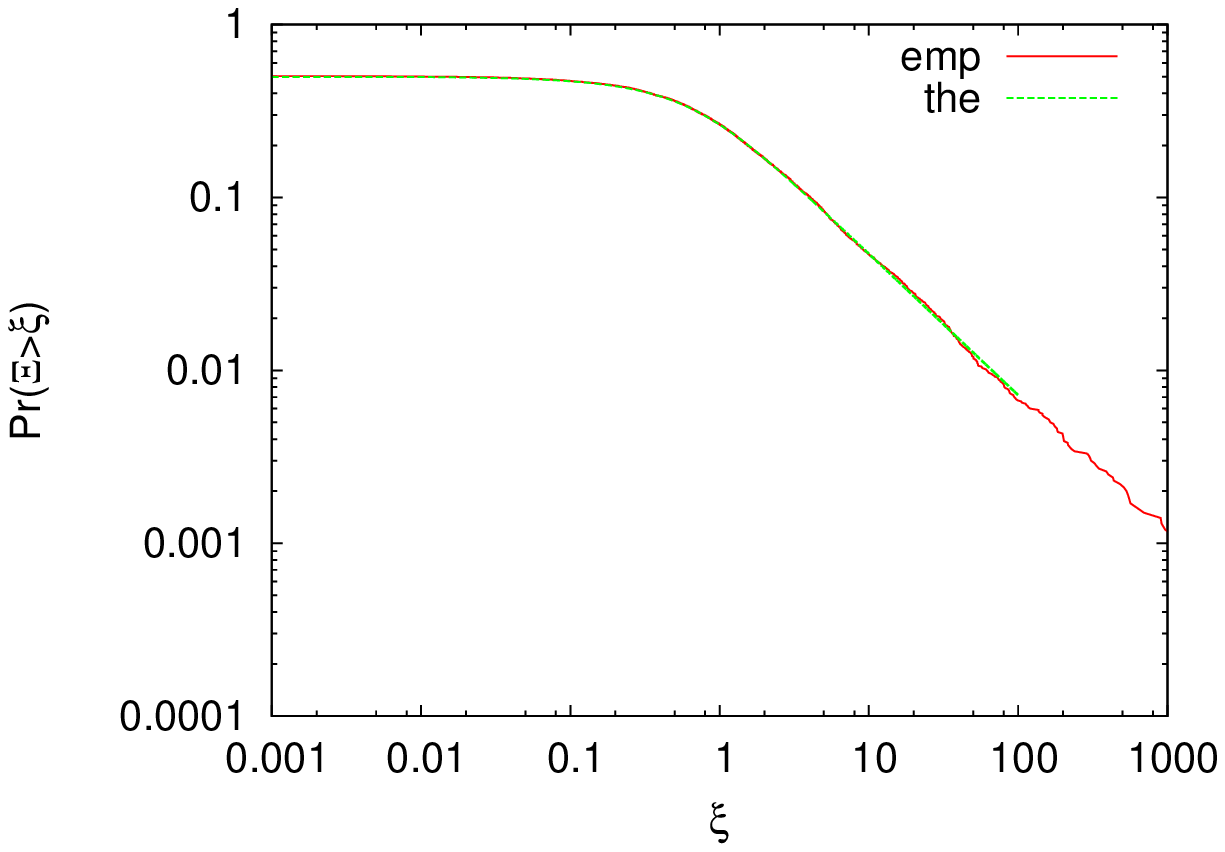}(g)
\caption{Complementary cumulative distribution functions of $\xi$ at $d=8$,
$l=2$, and $c=1$ for (a) $q'=-0.9$, (b) $-0.4$, (c) $0.1$, (d) $0.6$, (e) $1.1$
($\textcolor{black}{\nu}=19$), (f) $1.6$ ($\textcolor{black}{\nu}=2.33$),
 and (g) $2.1$ ($\textcolor{black}{\nu}=0.818$).
 Red curves represent empirical distributions, 
 and green ones represent theoretical distributions.}
\label{fig:cdf}
\end{figure}

The cumulative distribution of $\xi$ generated by
Equation \ref{eq:circle}. Equation \ref{eq:z-map}. and Equation \ref{eq:map}, defined as 
\begin{equation}
\mbox{Pr}(\Xi \leq \xi) = \int_{-\infty}^{\xi}p_{\Xi}(\xi')\mbox{d}\xi',
\end{equation}
can be expressed as
\textcolor{black}{
\begin{equation}
\nonumber
\mbox{Pr}(\Xi \leq \xi) = 
\nonumber
\left\{
\begin{array}{ll}
\left\{
\begin{array}{ll}
1 & \xi > \sqrt{\frac{3-q'}{1-q'}} \\
 \frac{1}{2}\Bigl[1+\mbox{sign}(\xi)\beta\bigl(\frac{1-q'}{3-q'}\xi^2;\frac{1}{2},\frac{2-q'}{1-q'}\bigr)\Bigr] &
|\xi|\leq\sqrt{\frac{3-q'}{1-q'}}  \\
0 & \xi < -\sqrt{\frac{3-q'}{1-q'}} 
\end{array}
\right. &  \mbox{for} \quad q < 1 \\
1-\frac{1}{2}\mbox{erfc}\Bigl(\frac{\xi}{\sqrt{2}}\Bigr) & \mbox{for}
\quad q = 1 \\
 \frac{1}{2}\Bigl[1+\mbox{sign}(\xi)\beta\bigl(\frac{\frac{q'-1}{3-q'}\xi^2}{1+\frac{q'-1}{(3-q')}\xi^2};\frac{1}{2},\frac{1}{q'-1}-\frac{1}{2}\bigr)\Bigr]
  & \mbox{for} \quad 1 < q < 3
\end{array}
\right., \\
\label{eq:theoretical-cumulative}
\end{equation}
}
where $\beta(x;a,b)$ is the regularized incomplete beta function,
\begin{equation}
\beta(x;a,b) = \frac{1}{B(a,b)}\int_{0}^{x}t^{a-1}(1-t)^{b-1}\mbox{d}t,
\end{equation}
and $\mbox{erfc}(x)$ is the complementary error function defined as
\begin{equation}
\mbox{erfc}(x) = \frac{2}{\sqrt{\pi}}\int_x^{\infty} e^{-t^2}\mbox{d}t.
\end{equation}
We compare the cumulative distributions of $\xi_n$ obtained from
Equation \ref{eq:circle}. Equation \ref{eq:z-map}, and Equation \ref{eq:map} 
with Equation \ref{eq:theoretical-cumulative}. Since we normally
generate $q$-Gaussian random variables from the given $q'$, for practical
usage, we need the inverse relation between $q$ and $q'$: $q = (q'+1)/(3-q')$.
Figure \ref{fig:cdf} shows the empirical complementary cumulative
distributions of $\xi$,
\begin{equation}
\mbox{Pr}(\Xi \geq \xi) = 1-\mbox{Pr}(\Xi \leq \xi),
\end{equation}
computed from 10,000 samples for $(w_0, z_0)=(0.1,1.0)$. Comparing
the empirical distribution with the theoretical one, we found that they
are very close for each parameter $q'$.

\textcolor{black}{
\begin{table}[ht]
\caption{\textcolor{black}{The best KS and AD statistics obtained from 10,000
  samples in 100 trials} for several $q$ at $d=8$, $l=2$, and $c=1$. 
$p$-values of both KS and AD tests  are shown.}
\label{tab:KStest8_1_2}
\centering
\begin{tabular}{llllll}
\hline
$q$ & $\textcolor{black}{\nu}$ & $p$-value (AD) & $p$-value (KS) \\
\hline
-1.0 & - & 0.996000 & 0.985991 \\
-0.9 & - & 0.994000 & 0.974883 \\
-0.8 & - & 0.997200 & 0.999401 \\
-0.7 & - & 0.995800 & 0.996095 \\
-0.6 & - & 0.995800 & 0.990724 \\
-0.5 & - & 0.992200 & 0.994215 \\
-0.4 & - & 0.992200 & 0.998253 \\
-0.3 & - & 1.000000 & 0.998473 \\
-0.2 & - & 0.995600 & 0.999262 \\
-0.1 & - & 1.000000 & 0.994604 \\
0.0 & - & 0.994400 & 0.988120 \\
0.1 & - & 0.996000 & 0.986996 \\
0.2 & - & 0.995800 & 0.998657 \\
0.3 & - & 0.996600 & 0.972872 \\
0.4 & - & 0.994800 & 0.979354 \\
0.5 & - & 0.995000 & 0.980051 \\
0.6 & - & 0.996600 & 0.993282 \\
0.7 & - & 0.997200 & 0.996990 \\
0.8 & - & 0.996000 & 0.984822 \\
0.9 & - & 0.996000 & 0.999201 \\
1.0 & - & 0.990400 & 0.998745 \\
1.1 & 19 & 0.992000 & 0.997347 \\
1.2 & 9 & 0.984400 & 0.963889 \\
1.3 & 5.66 & 0.994400 & 0.983520 \\ 
1.4 & 4 & 0.996400 & 0.995607 \\ 
1.5 & 3 & 0.994400 & 0.997567 \\ 
1.6 & 2.33 & 0.995000 & 0.999408 \\ 
1.7 & 1.85 & 0.996200 & 0.990171 \\ 
1.8 & 1.5 & 0.995600 & 0.994572 \\ 
1.9 & 1.22 & 0.996800 & 0.984163 \\ 
2.0 & 1 & 0.995800 & 0.995830 \\ 
2.1 & 0.818 & 0.992000 & 0.995684 \\  
2.2 & 0.666 & 0.996000 & 0.982779 \\ 
2.3 & 0.538 & 0.995800 & 0.993096 \\ 
2.4 & 0.428 & 0.000000 & 0.875352 \\ 
2.5 & 0.333 & 0.000000 & 0.995347 \\ 
2.6 & 0.25 & 0.000000 & 0.994969 \\ 
2.7 & 0.176 & 0.000000 & 0.007562 \\ 
2.8 & 0.111 & 0.000000 & 0.000000 \\ 
2.9 & 0.052 & 0.000000 & 0.000000 \\ 
\hline
\end{tabular}
\end{table}
}

\textcolor{black}{
\begin{table}[ht]
\caption{\textcolor{black}{The best KS and AD statistics obtained from 10,000
  samples in 100 trials} for several $q$ at $d=6$, $l=2$, and $c=6$. $p$-values of both KS and AD tests  are shown.} 
\label{tab:KStest6_6_2}
\centering
\begin{tabular}{llllll}
\hline
$q$ & $\textcolor{black}{\nu}$ & $p$-value (AD) & $p$-value (KS) \\
\hline
-1.0 & - & 0.993600 & 0.998321 \\
-0.9 & - & 0.994800 & 0.999067 \\
-0.8 & - & 0.994400 & 0.997075 \\
-0.7 & - & 0.996200 & 0.994040 \\
-0.6 & - & 0.991600 & 0.992699 \\
-0.5 & - & 0.994600 & 0.999246 \\
-0.4 & - & 0.995000 & 0.973464 \\
-0.3 & - & 0.995400 & 0.992854 \\
-0.2 & - & 0.995600 & 0.983395 \\
-0.1 & - & 0.994800 & 0.992643 \\
0.0 & - & 0.997000 & 0.980508 \\
0.1 & - & 0.995800 & 0.996620 \\
0.2 & - & 0.996000 & 0.999265 \\
0.3 & - & 0.996600 & 0.970387 \\
0.4 & - & 0.996200 & 0.992929 \\
0.5 & - & 0.996000 & 0.999219 \\
0.6 & - & 0.992200 & 0.995459 \\
0.7 & - & 0.996400 & 0.991304 \\
0.8 & - & 0.994400 & 0.959594 \\
0.9 & - & 0.995800 & 0.999786 \\
1.0 & - & 0.993800 & 0.997754 \\
1.1 & 19 & 0.996400 & 0.998304 \\
1.2 & 9 & 0.979800 & 0.959894 \\
1.3 & 5.66 & 0.99600 & 0.999363 \\
1.4 & 4 & 0.995800 & 0.987967 \\
1.5 & 3 & 0.994800 & 0.978924 \\
1.6 & 2.33 & 0.995800 & 0.999754 \\
1.7 & 1.85 & 0.996400 & 0.994942 \\
1.8 & 1.5 & 0.995400 & 0.999325 \\
1.9 & 1.22 & 0.997000 & 0.994694 \\
2.0 & 1 & 0.996400 & 0.978461 \\
2.1 & 0.818 & 0.988800 & 0.999509 \\
2.2 & 0.666 & 0.996400 & 0.991371 \\
2.3 & 0.538 & 0.996400 & 0.997778 \\
2.4 & 0.428 & 0.000000 & 0.928166 \\
2.5 & 0.333 & 0.000000 & 0.981747 \\
2.6 & 0.25 & 0.000000 & 0.989397 \\
2.7 & 0.176 & 0.000000 & 0.007562 \\
2.8 & 0.111 & 0.000000 & 0.000000 \\
2.9 & 0.052 & 0.000000 & 0.000000 \\
\hline
\end{tabular}
\end{table}
}

\textcolor{black}{We conducted the Kolmogorov-Smironov (KS) and
  the Anderson-Darling (AD) tests in order to verify whether the empirical
  distributions of sequences generated by our proposed method are
  convergent to the $q$-Gaussian distributions.} \textcolor{black}{It
  is known that Anderson--Darling test is suitable for checking the
  goodness-of-fit for heavy-tailed distributions~\cite{Anderson}.}
\textcolor{black}{Assuming $M$ samples of $\xi_1, \ldots, \xi_M$, the
  test statistics are given as
\begin{equation}
Z = \sqrt{M}\max_{n}\Bigl|F_M(\xi_n)-\mbox{Pr}(\Xi \geq
\xi_n)\Bigr|\sqrt{\psi\bigl(\mbox{Pr}(\Xi \geq \xi_n)\bigr)},
\end{equation}
where $F_M(\xi_n)$ an empirical cumulative distribution function, and $\psi(u)$
is a weight function. In the case of $\psi(u) = 1$, $Z$ gives a KS
test statistic and in the case of $\psi(u)=\frac{1}{u(1-u)}$, $Z$
gives an AD test statistic.}

Table \ref{tab:KStest8_1_2} shows the \textcolor{black}{best} $p$-values 
of both KS and AD tests for several $q$ values \textcolor{black}{at
$d=8$, $l=2$, and $c=1$.} The $p$-value of KS test is greater than
0.1 for $q < 2.7$. Therefore, the null hypothesis that the sequences are not 
samples from the theoretical distribution is not rejected at more than
5\% statistical significance for $q$ values from 1 to 2.6 in KS test. 
The degree of freedom $\nu$ goes to 0 as $q$ approaches 3. 
\textcolor{black}{For $q > 2.7$ ($\textcolor{black}{\nu} < 0.17$), both
the proposed procedure and GBMM does not work since degree of freedom $\nu$
is very small. The $p$-value of AD test is greater than 0.1
for $q < 2.4$. Since AD test is sensitive for tail events, the null
hypothesis is not rejected from the value of $q$ smaller than KS test
values. Table \ref{tab:KStest6_6_2} shows the $p$-values of both KS
and AD tests for several $q$ values at $d=6$, $l=2$, and $c=6$. The
tendency of $p$-values is very similar to ones at $d=8$, $l=2$, and 
$c=1$. The KS test passes at more than 5\% statistical
significance for $q$ values ranging from -1 to 2.6 in KS test. The
same is true for $-1 \leq q < 2.4$ in the case of AD test.}

\begin{figure}[!t]
\centering
\includegraphics[scale=0.55]{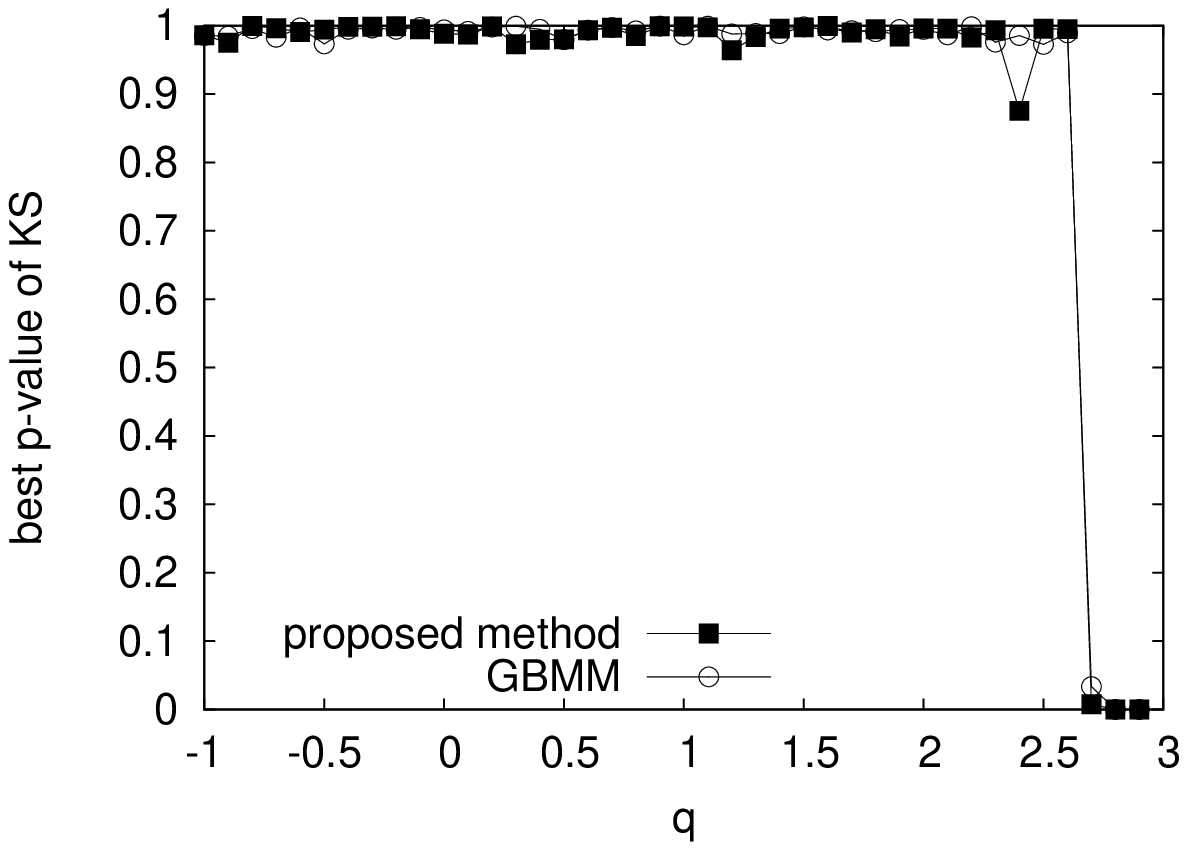}(a)
\includegraphics[scale=0.55]{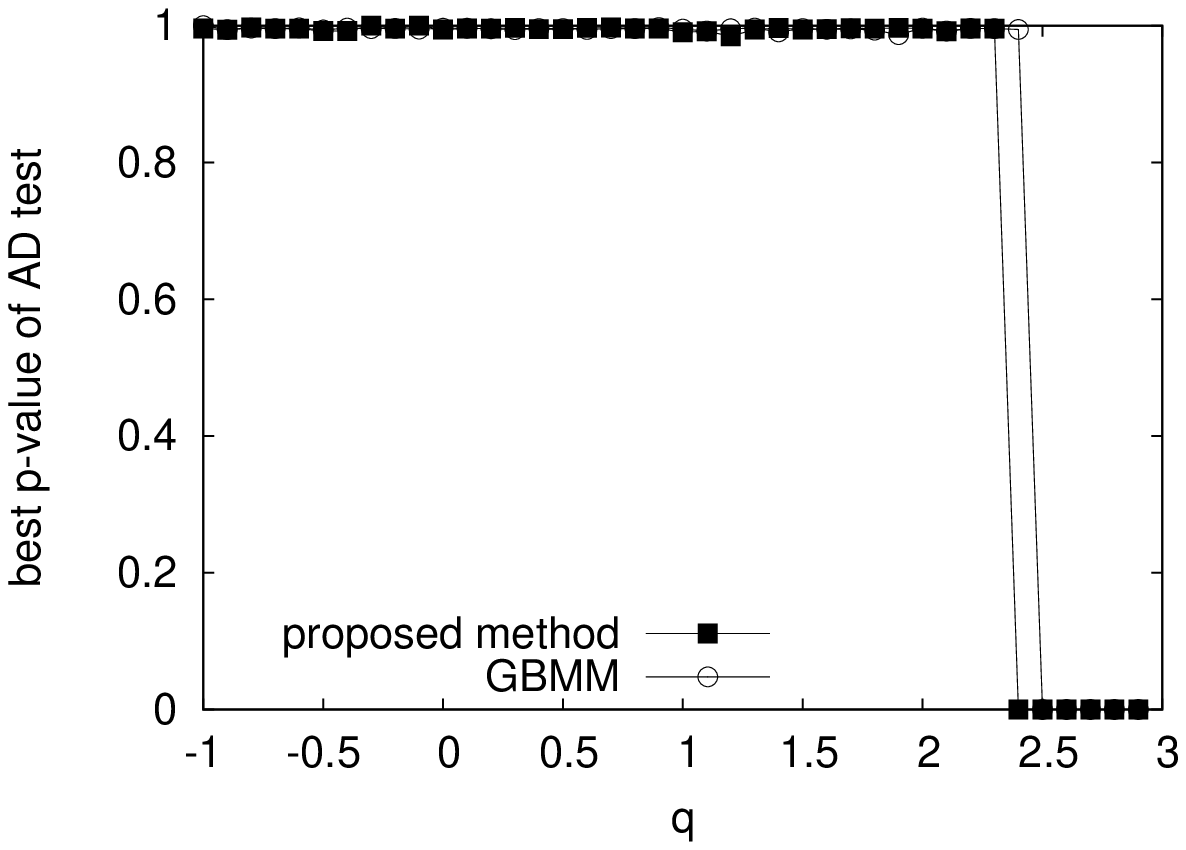}(b)
\caption{\textcolor{black}{(a) The best $p$-values of both (a) KS and
    (b) AD tests obtained from 10,000 samples in 100 trials with our
    proposed and GBMM for several $q$ at $d=8$, $l=2$, and $c=1$.}}
\label{fig:comp8_1_2}
\end{figure}

\begin{figure}[!t]
\centering
\includegraphics[scale=0.55]{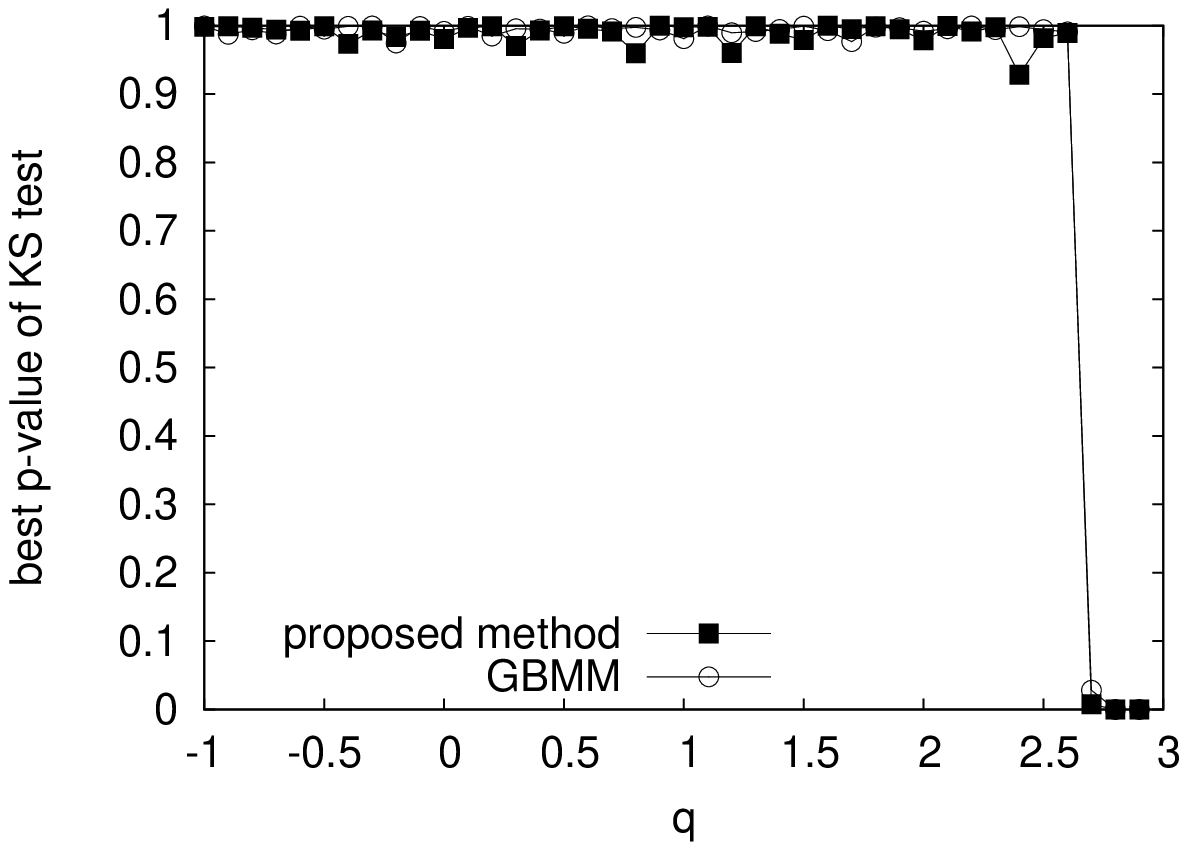}(a)
\includegraphics[scale=0.55]{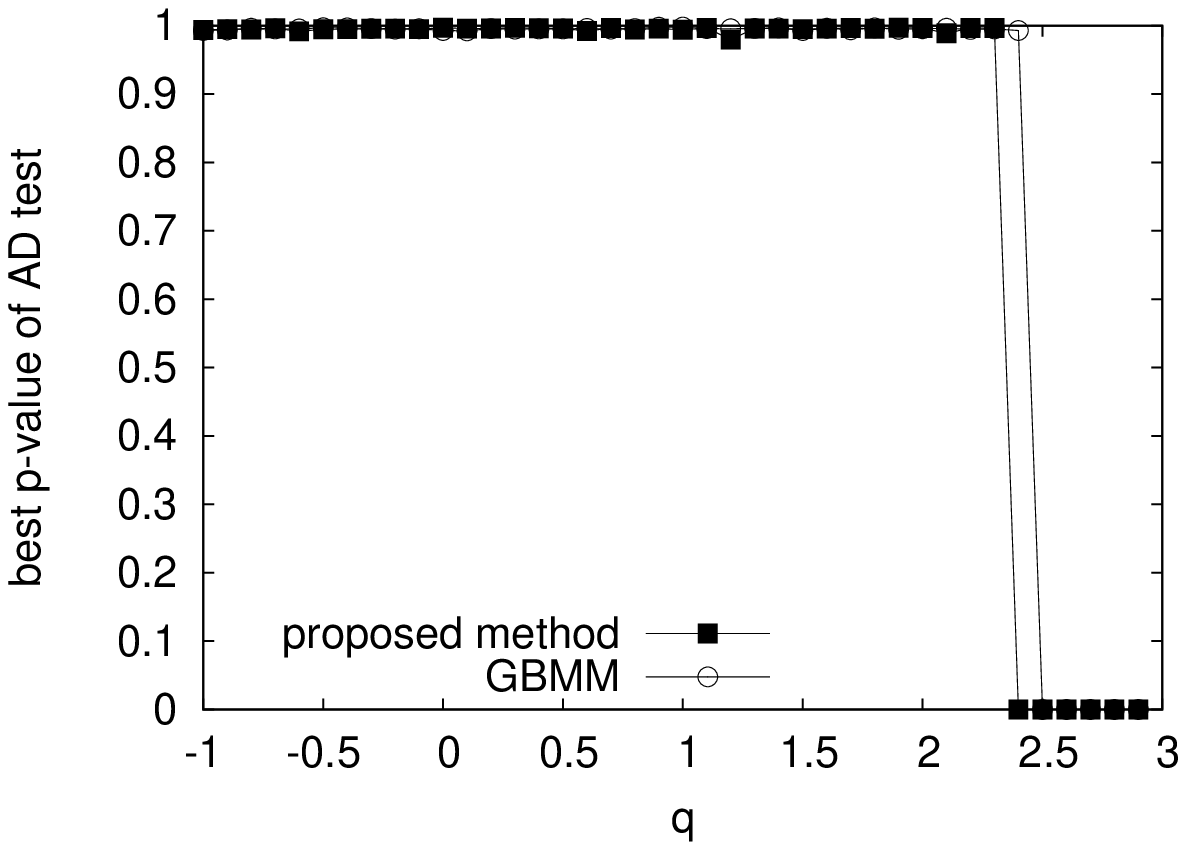}(b)
\caption{\textcolor{black}{(a) The best $p$-values of both (a) KS and
    (b) AD tests obtained from 10,000 samples in 100 trials with our
    proposed and GBMM for several $q$ at $d=6$, $l=2$, and $c=6$. }}
\label{fig:comp6_6_2}
\end{figure}

\textcolor{black}{While} GBMM~\cite{Thistleton:07} \textcolor{black}{is}
based on transformation of uniform random variables\textcolor{black}{, our}
\textcolor{black}{proposed} method \textcolor{black}{here}  
is purely mechanical generation of $q$-Gaussian distribution
\textcolor{black}{based on ergodic theory}. \textcolor{black}{Thus,
  no random number are not assumed for the generations of $q$-Gaussian
  distribution.} Its implementation is very simple as shown  
in \textcolor{black}{the example code in} Appendix
~\ref{sec:source}\textcolor{black}{.} \textcolor{black}{Figures 
\ref{fig:comp6_6_2} ($d=8$, $l=2$, and $c=1$) and 
\ref{fig:comp8_1_2} ($d=6$, $l=2$, and $c=6$) show the best p-values 
of (a) KS test and (b) AD test obtained from 10,000 samples in 100 trials 
with the proposal and the GBMM for several $q$. The best $p$-values 
provided by the proposed method are same as ones by the GBMM for many cases.}

\section{Conclusion}
We proposed a pseudo random \textcolor{black}{number} generator of $q$-Gaussian
random variables for a range of $q$ values, $-\infty < q < 3$,
based on deterministic map dynamics. Our method consists of
ergodic transformation on the unit circle and map dynamics based 
on the piecewise linear map. We conducted \textcolor{black}{both
KS and AD tests for random number sequences generated by
GBMM and our proposed chaotic method for several values of $q$.} 
The $q$-Gaussian samples passed the KS test at the 5\% significance level 
for $q < 2.7$, and passed the AD test at the 5\%significance level for 
$q < 2.4$.

\appendices
\section{Source code}
\label{sec:source}
We show a C source code for our proposed method \textcolor{black}{for
$d=8$, $l=2$, and $c=1$}. The code is exhibited in order to
demonstrate the algorithm, and is not optimal for speed. The algorithm
is implemented in four functions. The first two functions compute
$q$-exponential and $q$-logarithmic functions. The next function
setseed\_qnormal($v_0$, $z_0$) sets two random seeds $v_0$ and $z_0$,
and qnormal($q$) calls the iterated map to generate $q$-Gauss random
variables by our proposed method.
{\small 
\begin{verbatim}
#include <stdio.h>
#include <math.h>
#include <stdlib.h>
#include <strings.h>
double qnormal_x,qnormal_y,qnormal_z;
double expq(double q, double w){
  if(q==1.0){
    return(exp(w));
  }
  else{
    return (expl(log(1.0+(1.0-q)*w)/(1.0-q)));
  }
}
double lnq(double q, double w){
  if(q==1.0){
    return(log(w));
  }
  else{
    return ((exp(log(w)*(1.0-q))-1.0)/(1.0-q));
  }
}
void setseed_qnormal(double v0, double z0){
  qnormal_x = sqrt(1-v0*v0); 
  qnormal_y = v0;
  qnormal_z = z0;
}
double Q8(double w, double v){
  return(8*w*v*(((16.0*w*w-24.0)*w*w+10.0)*w*w-1.0));
}
double P8(double w){
  return((((128.0*w*w-256.0)*w*w+160.0)*w*w-32.0)*w*w+1.0);
}
double f(double z){
  return(1.0-fabs(1.0-1.99999*z));
}
void qnormal(double q){
  double qq;
  qnormal_y = Q8(qnormal_x,qnormal_y);
  qnormal_x = P8(qnormal_x);
  qq = (q+1.0)/(3.0-q); 
  qnormal_z = f(expq(qq,-qnormal_z*qnormal_z*0.5));
  qnormal_z = sqrt(-2.0*lnq(qq,qnormal_z));
}
int main(int argc, char *argv[]){
  double q,v0,z0,eta,xi;
  int i;
  if(argc != 4){
     printf("%s q v0 z0\n",argv[0]);
     exit(0);
  }
  q = (double)atof(argv[1]);
  v0 = (double)atof(argv[2]);
  z0 = (double)atof(argv[3]);
  setseed_qnormal(v0,z0);
  for(i=0;i<10000;i++){
    qnormal(q);
    xi = qnormal_x*qnormal_z;
    eta = qnormal_y*qnormal_z;
    printf("%lf %lf\n",xi,eta);
  }
}
\end{verbatim}
}



\ifCLASSOPTIONcaptionsoff
  \newpage
\fi

\end{document}